\documentclass[10pt]{article}

\usepackage{graphicx} 
\usepackage{cite}     
\usepackage{amsthm}
\usepackage{amsmath}
\usepackage{amssymb}
\usepackage{amsfonts}

\newcommand{\intf}{{\mathrm{int}}}
\newcommand{\dd}{{\rm{d}}} 
\newcommand{\boldu}{\mbox{\boldmath$u$}} 
\newcommand{\bolde}{\mbox{\boldmath$e$}} 
\newcommand{\boldk}{\mbox{\boldmath$k$}} 
\newcommand{\boldl}{\mbox{\boldmath$l$}} 
\newcommand{\boldm}{\mbox{\boldmath$m$}} 
\newcommand{\cH}{{\mathcal{H}}}
\newcommand{\cR}{{\mathcal{R}}}

\def \sR {{{\,}^S\! R}{}}
\def \sH {{{\,}^S\! H}{}}
\def \sL {{L_{GBT}}{}}
\def \Spq {Q_{pq}}
\def \Spn {Q_{pn}}
\def \Scontramn {Q^{mn}}
\def \Scontrapq {Q^{pq}}

\begin{document}

\title{Kundt spacetimes in the Einstein--Gauss--Bonnet theory}

\author{R.~\v{S}varc, J.~Podolsk\'y, O. Hru{\v s}ka \\
\vspace{0.05cm} \\
\\
\vspace{0.05cm} \\
{\small Institute of Theoretical Physics, Charles University, Prague,} \\
{\small Faculty of Mathematics and Physics, V~Hole\v{s}ovi\v{c}k\'ach~2, 180~00 Praha 8, Czech Republic.} \\[3mm]
{\small E-mail: \texttt{
robert.svarc@mff.cuni.cz,
podolsky@mbox.troja.mff.cuni.cz,
hruskaondrej@seznam.cz}}\\
}

\maketitle

\begin{abstract}
We systematically investigate the complete class of vacuum solutions in the Einstein--Gauss--Bonnet gravity theory which belong to the Kundt family of non-expanding, shear-free and twist-free geometries (without gyratonic matter terms) in any dimension. The field equations are explicitly derived and simplified, and their solutions classified into three distinct subfamilies. Algebraic structures of the Weyl and Ricci curvature tensors are determined. The corresponding curvature scalars directly enter the invariant form of equation of geodesic deviation, enabling us to understand the specific local physical properties of the gravitational field constrained by the EGB theory. We also present and analyze several interesting explicit classes of such vacuum solutions, namely the Ricci type III spacetimes, all geometries with constant-curvature transverse space, and the whole {\it pp\,}-wave class admitting a covariantly constant null vector field. These exact Kundt EGB gravitational waves exhibit new features which are not possible in Einstein's general relativity.
\end{abstract}

\vfil\noindent
PACS class: 04.20.Jb, 04.50.-h, 04.50.Kd, 04.30.-w, 04.30.Nk




\section{Introduction}
\label{Sec:Intro}

The Kundt spacetimes, introduced in \cite{Kundt:1961,Kundt:1962}, represent one of the most impressive classes of exact solutions within the classic Einstein's general relativity, as well as its higher-dimensional extensions \cite{PodolskyZofka:2009, ColeyEtal:2009}. Their notable particular members such as {\it pp\,}-waves, VSI spacetimes,\footnote{Defined as geometries for which all the curvature scalar invariants vanish.} or direct-product spacetimes have become textbook models providing a deeper insight into the structure of the Einstein gravity theory and its various inherent properties, see comprehensive monographs \cite{Stephanietal:2003, GriffithsPodolsky:2009}. Interestingly, in an arbitrary dimension the Kundt family is defined invariantly in terms of \emph{optical scalars} as those geometries admitting \emph{non-twisting}, \emph{shear-free}, and \emph{non-expanding} null geodesic congruence, see e.g. \cite{OrtaggioPravdaPravdova:2013, PodolskySvarc:2013a} for the review and detailed list of references. This purely geometric definition thus holds irrespectively of a specific metric field theory of gravity. However, particular field equations of a given gravity theory put further specific restrictions on the resulting spacetime. The Kundt class thus provides a unique non-trivial opportunity to \emph{compare distinct theories of gravity} on the level of the corresponding exact solutions.

In the coordinate setting, which is naturally adapted to its geometry, the $D$-dimensional Kundt manifold is described by the line element
\begin{align}
\mathrm{d} s^2=g_{pq}(u,x)\,\dd x^p\,\dd x^q+2g_{up}(r,u,x)\,\dd u\,\dd x^p -2\,\dd u\,\dd r
+g_{uu}(r,u,x)\,\dd u^2 \,, \label{Kundt_metric}
\end{align}
where $r$ represents the \emph{affine parameter} along null geodesics forming the non-twisting, shear-free, and non-expanding congruence  generated by the vector field $\boldk$ (i.e. ${\boldk=\partial_r}$), the coordinate $u$ labels \emph{null hypersurfaces} with $\boldk$ normal (and also tangent) whose existence is guaranteed by the Poincare lemma, and ${x^p}$ with $p$ ranging from $2$ to ${D-1}$ cover the Riemannian \emph{transverse space} with $u$ and $r$ fixed. An important attribute of the Kundt class is the $r$-independence of the corresponding transverse metric ${g_{pq}}$ (which is in contrast to the expanding Robinson--Trautman class \cite{PodolskySvarc:2015a}). Due to the gauge freedom of the line element (\ref{Kundt_metric}), see \cite{PodolskyZofka:2009, ColeyEtal:2009}, the off-diagonal metric functions ${g_{up}}$ can be simplified or even completely removed (at least locally). The exceptional case, for which these terms carry physical information, corresponds to so-called gyratonic solutions representing a beam of null radiation with internal spin \cite{Bonnor:1970b, FrolovFursaev:2005, KrtousPodolskyZelnikovKadlecova:2012, PodolskySteinbauerSvarc:2014}. The focus of this paper are generic vacuum spacetimes without internal angular momentum, so that we set
\begin{equation}
g_{up}=0 \,. \label{non-gyr_Condition}
\end{equation}
Such most general non-gyratonic \emph{Kundt geometries} take the form
\begin{equation}
\dd s^2 = g_{pq}(u,x)\, \dd x^p\,\dd x^q -2\,\dd u\,\dd r+g_{uu}(r,u,x)\, \dd u^2 \,, \label{Kundt_non_gyr}
\end{equation}
with non-trivial contravariant metric components given by
\begin{equation}
g^{pq}\,, \qquad g^{ru}=-1\,, \qquad g^{rr}= -g_{uu} \,, \qquad \hbox{where} \qquad {g_{pk}\,g^{kq}={\delta_p}^q} \,. \label{Non_triv_contravar_met}
\end{equation}

Throughout this paper, we employ the \emph{Einstein--Guauss--Bonnet} gravity (EGB) to restrict the (non-gyratonic) general Kundt line element (\ref{Kundt_non_gyr}). This famous theory arises as the simplest non-trivial representative of a large class of Lovelock gravities \cite{Lovelock:1971} or also, for example, as the limit of the heterotic string theory \cite{Gross:1987, Bento:1996} for low energies. Its vacuum action in $D\geq5$ dimensions is given by
\begin{align}\label{action}
S=\int\!\left[{\kappa}^{-1}\left(R-2\Lambda_0\right) +\gamma\,L_{GB}\right]\sqrt{-g}\,\,\dd^D \rm{x} \,,
\end{align}
where $R$ is the Ricci scalar, ${\Lambda_0}$, ${\kappa}$ and ${\gamma}$ are the theory constants, and ${L_{GB}}$ represents the Gauss--Bonnet term
\begin{equation}
L_{GB}\equiv R^2_{cdef}-4\,R^2_{cd}+R^2 \,, \label{GB_term}
\end{equation}
constructed as specific combination of the scalar curvature squares, namely $R^2$,
\begin{equation}
R^2_{cdef}\equiv R_{cdef}\,R^{cdef} \qquad\hbox{and}\qquad
R^2_{cd}\equiv R_{cd}\,R^{cd} \,.
 \label{def_quatratic_terms}
\end{equation}
The field equations induced by the action (\ref{action}) read
\begin{align}\label{GB_Field_eqns}
\frac{1}{\kappa}\left(R_{ab}-\tfrac{1}{2}R\,g_{ab}+\Lambda_0\,g_{ab}\right)+2\gamma\,H_{ab} =0 \,,
\end{align}
where $H_{ab}$ stands for
\begin{equation}\label{GB_FE_contrib}
H_{ab}\equiv R\,R_{ab}-2R_{acbd}\,R^{cd}+R_{acde}\,R_{b}{}^{cde}-2R_{ac}\,R_{b}{}^{c}
-\tfrac{1}{4}\,g_{ab}\,L_{GB}\,.
\end{equation}
It is also useful to express their trace
\begin{equation}\label{TraceFieldEqs}
R = \frac{2}{D-2}\,\big[D\Lambda_0+2\kappa\gamma\, H\big]\, , \qquad \hbox{with} \qquad H\equiv g^{ab}H_{ab}=-\frac{1}{4}(D-4)L_{GB} \,,
\end{equation}
and then rewrite the field equations (\ref{GB_Field_eqns}) as
\begin{equation}
R_{ab}=\frac{2\Lambda_0}{D-2}g_{ab}-2k\left(H_{ab}-\frac{g_{ab}}{D-2}H\right), \qquad \hbox{where} \qquad {k\equiv\kappa\gamma}\,. \label{RicciFieldEqs}
\end{equation}
Our main aim here is to explicitly derive and analyze these 2nd-order field equations for spacetimes of the form (\ref{Kundt_non_gyr}). Of course, for ${\gamma=0=k}$ the system~(\ref{GB_Field_eqns}) reduces to classic Einstein's equations. We can thus directly compare mathematical and physical properties of obtained solutions in the Einstein--Guauss--Bonnet gravity with those studied for more than half a century in the framework of Einstein's general relativity. Some particular results related to the Kundt geometries (\ref{Kundt_non_gyr}) have been already presented in our previous works \cite{OH:2016, MK:2017}, but here we proceed in full generality, supplemented by a deeper geometric and physical analysis. Moreover, some complementary results obtained in the context of general Lovelock gravity can be found in \cite{MO:2018}.

Finally notice that recently a specific approach was suggested in \cite{GlavanLin:2020} to introduce the EGB theory even in standard dimension ${D=4}$. Immediately, dozens of specific applications, see e.g., \cite{KZ:2020, GM:2020}, have followed, together with some doubts about the physical relevance of this method, see e.g., \cite{GST:2020, HKMP:2020}. Comprehensive list of the related references can be found e.g. in \cite{LM:2020}. Even though our calculations here are fully general, and the non-trivial particular limit ${D\rightarrow 4}$ for the Kundt geometries can be, in principal, obtained, such analysis goes beyond the scope of this work and will be presented elsewhere. Here, let us only remark that the key quantity ${H_{ab}}$ is not in the case of general transverse metric ${g_{pq}}$ factorized by ${(D-4)}$ which may lead to the singular behaviour in a combination with the redefinition of the theory parameter ${k \to {k}/{(D-4)}}$.

The paper is organized as follows. In Section~\ref{Sec:FEqs} we formulate the field equations, employ their constraints, derive the general solution, and distinguish particular distinct cases. To discuss the physically relevant properties of new spacetimes, we review algebraic structure of the curvature tensors in Section~\ref{Sec:AlgStr} which we subsequently use to study the geodesic deviation in a coordinate-independent form in Section~\ref{Sec:GeodDev}. These tools are then employed in Section~\ref{Sec:Subclasses} to analyze the most interesting representatives of the Kundt class, and compare the Einstein and Einstein--Guauss--Bonnet theories. Finally, in Appendices~\ref{appendixA} and \ref{appendixB} the curvature tensors for the metric (\ref{Kundt_non_gyr}) and their quadratic contractions used in this paper are listed, respectively.


\section{Einstein--Gauss--Bonnet field equations for the Kundt class and their systematic solution}
\label{Sec:FEqs}

To derive the complete family of Kundt solutions (\ref{Kundt_non_gyr}) in the Einstein--Gauss--Bonnet gravity, we first calculate all necessary coordinate components of the curvature tensor and their combinations which appear in the field equations~(\ref{GB_Field_eqns}), (\ref{GB_FE_contrib}). These quantities are summarized in Appendix~\ref{appendixA} and \ref{appendixB}, respectively.

Since the metric functions ${g_{rr}}$ and ${g_{rp}}$ are zero, and also the ${rr}$- and ${rp}$-components of the relevant tensor contractions vanish, we observe that these components of the field equations~(\ref{GB_Field_eqns}) are satisfied \emph{identically}, both in the Einstein as well as the Einstein--Gauss--Bonnet theories. It remains to investigate the non-trivial components, namely ${ru}$, ${pq}$, ${up}$, and ${uu}$ to restrict the metric functions in (\ref{Kundt_non_gyr}).

\begin{itemize}

\item
The ${ru}$-component of the field equations~(\ref{GB_Field_eqns}) connects the geometry of the ${(D-2)}$-dimensional transverse space, described by the Riemannian metric $g_{pq}(u,x)$, to the constant parameters ${\Lambda_0}$, $k$ of the theory, namely
\begin{align}
& \sR-2\Lambda_0+k\left(\sR^2_{klmn}-4\sR^2_{mn}+\sR^2\right)=0 \,. \label{GB_ru}
\end{align}
The curvature quantities with the superscript ${\,^S}$ are calculated with respect to the spatial metric $g_{pq}$. In the case of classic general relativity (${k=0}$), we immediately obtain that the transverse-space Ricci scalar curvature ${\sR}$ has to be a constant equal to ${2\Lambda_0}$. This is no more true in the more general Einstein--Gauss--Bonnet theory, where it is also coupled to the Gauss--Bonnet term constructed from the transverse-space metric~$g_{pq}$.

\item
The ${pq}$-component of the field equations~(\ref{GB_Field_eqns}), combined with the algebraic constraint (\ref{GB_ru}), gives
\begin{align}
& \Spq \, g_{uu,rr} +\sR_{pq} \nonumber\\
& +2k\left(\sR_{pq}\sR-2\sR_{pmqn}\sR{}^{mn}+\sR_{pklm}\sR_q{}^{klm}-2\sR_{pm}\sR_q{}^m  \right) = 0 \,, \label{GB_pq_s}
\end{align}
where ${\Spq}$ is a \emph{fundamental quantity} defined as
\begin{equation}
\Spq \equiv -\tfrac{1}{2}\,g_{pq}+k\left(2\sR_{pq}-\sR\,g_{pq}\right).
\label{Spq}
\end{equation}
Its trace is
\begin{equation}
Q \equiv g^{pq}\,\Spq
  = -\left[\tfrac{1}{2}(D-2)+k(D-4)\sR\right] .
\label{SpqTRACE}
\end{equation}

Evaluating the \emph{trace} of the field equation (\ref{GB_pq_s}), we obtain a simple explicit constraint
\begin{align}
-Q\,g_{uu,rr} = 4\Lambda_0 - \sR \,. \label{GB_tr_s}
\end{align}
In combination with (\ref{SpqTRACE}), after integration this determines the ${r}$-dependence of the metric function ${g_{uu}}$. Further discussion must be split into distinct cases, namely ${Q \ne 0}$ and ${Q=0}$.

\item
The ${up}$-component of the system~(\ref{GB_Field_eqns}), simplified by using previous equations (\ref{GB_ru}) and (\ref{GB_pq_s}), takes the form
\begin{align}
\Spn \,g^{nm}\big(&g_{uu,rm}-2g^{kl}g_{k[m,u||l]}\big)  \nonumber\\
&+2k\big(-2\sR{}^{kl}\,\delta_p^m+\sR_p{}^{kml}\big)g_{k[m,u||l]}=0 \,, \label{GB_up_s}
\end{align}
where ${\,_{||}}$ denotes the covariant derivative on the transverse Riemannian space of dimension ${(D-2)}$. This equation can be understood as the \emph{constraint on the spatial dependence} of ${g_{uu}}$, and also the admitted $u$-dependence of the spatial metric~${g_{pq}}$.

\item
Finally, the ${uu}$-component of the field equations~(\ref{GB_Field_eqns}) can be written as
\begin{align}
\Scontrapq \big(&g_{uu||pq}+g_{pq,uu}-\tfrac{1}{2}g_{uu,r}\,g_{pq,u}-\tfrac{1}{2}g^{kl}g_{kp,u}\,g_{lq,u}\big)\nonumber\\
&+2k\,\big(g^{ko}g^{ls}-2\,g^{kl}g^{os}\big)g^{pq}\,g_{k[p,u||l]}\,g_{o[q,u||s]}=0 \,, \label{GB_uu_s}
\end{align}
which \emph{restricts the amplitudes of the transverse gravitational waves} encoded in $g_{uu||pq}$, see Sec.~\ref{Sec:AlgStr}.

\end{itemize}

To summarize, the conditions (\ref{GB_ru}), (\ref{GB_pq_s}) [implying (\ref{GB_tr_s})], with (\ref{GB_up_s}) and (\ref{GB_uu_s}) are the explicit and compact form of the field equations~(\ref{GB_Field_eqns}) for the generic (non-gyratonic) Kundt line element (\ref{Kundt_non_gyr}).

In the following subsections~\ref{SubS:neq0},~\ref{SubS:0neq0}, and~\ref{SubS:00} we will discuss \emph{three distinct subclasses} of these spacetimes in the Einstein--Gauss--Bonnet gravity, depending to the quantity ${\Spq }$ and its trace $Q$, defined by (\ref{Spq}) and (\ref{SpqTRACE}). They differ according to ${Q \neq 0}$, ${Q = 0}$, and ${\Spq = 0}$.


\subsection{Case ${Q \neq 0}$}
\label{SubS:neq0}

In this general case, the equation (\ref{GB_tr_s}) with (\ref{SpqTRACE}) can be immediately integrated to obtain the $r$-dependence of the metric function ${g_{uu}}$, namely
\begin{align}
g_{uu}(r,u,x)=b(u,x)\,r^2+c(u,x)\,r+d(u,x) \,, \label{GB_g_uu}
\end{align}
where the coefficient of the leading (quadratic) term is explicitly given by
\begin{equation}
b=\frac{4\Lambda_0-\sR}{(D-2)+2k\,(D-4)\sR} \,, \label{GB_b_Explicit}
\end{equation}
and $c(u,x)$, $d(u,x)$ are arbitrary functions. Substituting the $g_{uu,rr}$ term back to the original ${pq}$-equation (\ref{GB_pq_s}), we obtain the relation
\begin{align}
&\big[(D-2)+4k(D-4)\big(1+k\sR\big)\sR+16k\Lambda_0 \big]\sR_{pq}-\big(1+2k\sR\big)\big(4\Lambda_0-\sR\big)g_{pq} \nonumber\\
&\ -2k\big[(D-2)+2k(D-4)\sR\big]\big(2\sR_{pmqn}\sR{}^{mn}-\sR_{pklm}\sR_q{}^{klm}+2\sR_{pm}\sR_q{}^m\big)=0 \,. \label{GB_pq-sgen}
\end{align}
This is the additional constraint to (\ref{GB_ru}) restricting the geometry of the ${(D-2)}$-dimensional transverse space in relation to the theory constants. For Einstein's gravity theory, it reduces to ${{(D-2)}\sR_{pq}=\left(4\Lambda_0-\sR\right)g_{pq}}$, and (\ref{GB_ru}) simplifies to ${\sR=2\Lambda_0}$, so that
\begin{align}
&\sR_{pq}=\frac{2\Lambda_0}{D-2}\,g_{pq}  \,.
\end{align}
In standard general relativity the transverse space in Kundt vacuum spacetimes must be an Einstein space.

Using (\ref{GB_g_uu}), (\ref{GB_b_Explicit}), the $up$-component (\ref{GB_up_s}) of the field equations with ${g_{uu,rm}}$ now becomes
\begin{align}
&\Spn \,g^{nm}\big(2\,b_{,m}\,r+c_{,m}-2g^{kl}g_{k[m,u||l]}\big) +2k\big(-2\sR{}^{kl}\delta_p^m+\sR_p{}^{kml}\big)g_{k[m,u||l]}=0 \,. \label{GB_up_Spq_non_zero}
\end{align}
This equation has to be satisfied both for terms linear in $r$, and the $r$-independent part, respectively. The first constraint requires
\begin{align}
\Spn \,g^{nm}\,b_{,m}=0 \,. \label{GB_up_1}
\end{align}
Interestingly, this restriction \emph{is identically satisfied} as a consequence of the covariant divergence of equations (\ref{GB_ru}) and (\ref{GB_pq_s}) when the Bianchi identities and their contractions are employed. The $r$-independent part of (\ref{GB_up_Spq_non_zero}) implies
\begin{align}
\Spn \,g^{nm}\left(c_{,m}-2g^{kl}g_{k[m,u||l]}\right)+2k\left(-2\sR{}^{kl}\,\delta_p^m+\sR_p{}^{kml}\right)g_{k[m,u||l]}=0 \,. \label{GB_up_r0}
\end{align}
It determines the \emph{spatial dependence of a coefficient} $c(u,x)$ in the metric function ${g_{uu}}$, coupled to the $u$-dependence of the transverse space metric $g_{km}$.

Finally, substituting the form (\ref{GB_g_uu}) of ${g_{uu}}$ into the $uu$-component (\ref{GB_uu_s}) of the field equations, we obtain
\begin{align}
\Scontrapq \Big[\,b_{||pq}\,r^2+\big(c_{||pq}-b\,g_{pq,u}\big)\,r
& +d_{||pq}-\tfrac{1}{2}\,c\,g_{pq,u}+g_{pq,uu}-\tfrac{1}{2}g^{kl}g_{kp,u}\,g_{lq,u}\Big] \nonumber\\
& +2k\big(g^{mo}g^{ns}-2\,g^{mn}g^{os}\big)g^{pq}g_{m[p,u||n]}\,g_{o[q,u||s]}=0 \,. \label{GB_uu_Spq_non_zero}
\end{align}
The term \emph{quadratic} in $r$ gives the condition
\begin{align}
\Scontrapq \,b_{||pq}=0 \,, \label{GB_uu_r2}
\end{align}
which again \emph{is identically satisfied}. Indeed, it follows from (\ref{GB_up_1}) by rearranging indices, performing a covariant derivative, and applying the Leibniz rule that
${\Scontramn \,b_{||mn}+k\big(2\sR{}^{mn}{}_{||n}-\sR_{,n}\,g^{mn}\big)\,b_{,m}=0}$.
The term in the round brackets vanishes identically due to the contracted Bianchi identities.

The condition given by the \emph{linear} term in $r$ becomes
\begin{align}
&\Scontrapq \left(c_{||pq}-b\,g_{pq,u}\right)=0 \,.
\end{align}
%
%
This equation can further be simplified\footnote{Let us remark that the structure of the field equations in the EGB theory is very similar to those studied in various scenarios within the Kundt class in Einstein's theory, see for example footnote~8 of \cite{OrtaggioPravda:2016} or section IV.C. of \cite{PodolskySvarc:2019}. Typically, the parts of the ${up}$ and ${uu}$ field equations which are proportional to linear powers of $r$ are identically satisfied. However, in the case of (\ref{GB_uu_r}), due to its greater complexity, we have not yet been able to prove this conjecture.} by expressing the Laplace-like term ${\Scontramn c_{||mn}}$ as a covariant divergence of (\ref{GB_up_r0}) and substituting for $b$ from (\ref{GB_b_Explicit}) to obtain
\begin{align}
\frac{4\Lambda_0-\sR}{D-2+2k\left(D-4\right)\sR}\,\Scontramn g_{mn,u}
-2g^{kl}\Scontramn g_{k[m,u||l]||n} +2k\left(\sR^{nkml}-2\sR^{kl}g^{mn}\right)g_{k[m,u||l]||n} = 0 \,. \label{GB_uu_r}
\end{align}

The remaining part of equation (\ref{GB_uu_Spq_non_zero}), which is \emph{independent of $r$}, gives the constraint on the coefficient $d(u,x)$ in the metric function ${g_{uu}}$ of the form (\ref{GB_g_uu}), namely
\begin{align}
\Scontrapq \big(& d_{||pq}-\tfrac{1}{2}\,c\,g_{pq,u}+g_{pq,uu}-\tfrac{1}{2}g^{kl}g_{kp,u}\,g_{lq,u}\big)\nonumber\\
&+2k\big(g^{mo}g^{ns}-2\,g^{mn}g^{os}\big)g^{pq}g_{m[p,u||n]}\,g_{o[q,u||s]}=0 \,. \label{Spq_neq0_uu_abs}
\end{align}
This condition determines possible form of the Kundt gravitational waves, encoded by the amplitudes $d_{||pq}$.

To summarize: The field equations which must be satisfied are (\ref{GB_pq-sgen}) for $g_{pq}$, (\ref{GB_up_r0}) for $c$, and (\ref{Spq_neq0_uu_abs}) for $d$.


\subsection{Case ${Q = 0}$ with ${\Spq \neq 0}$}
\label{SubS:0neq0}

There may occur a peculiar situation in which ${\Spq \neq 0}$, but its trace vanishes. In such a case, ${Q = 0}$ implies a strict constraint on (\ref{SpqTRACE}) which uniquely fixes the transverse-space scalar curvature,
\begin{align}
\sR = -\frac{D-2}{2k(D-4)} \,, \label{GB_Ein_R}
\end{align}
which has to be non-vanishing and constant. This case is clearly \emph{not allowed in the Einstein theory}. Moreover, equation (\ref{GB_tr_s}) immediately implies
\begin{equation}
\sR=4\Lambda_0 \,.  \label{Rje4Lambda}
\end{equation}
Putting these two conditions together, we obtain the \emph{necessary coupling of all three theory parameters} as
\begin{align}
8(D-4)k\Lambda_0 = -(D-2) \,, \label{GBPar}
\end{align}
i.e., the relation
\begin{align}
\Lambda_0=-\frac{D-2}{8\,k\,(D-4)} \,. \label{GBParam}
\end{align}
For any Gauss--Bonnet parameter ${\gamma=k/\kappa}$ there is a unique value of the cosmological constant $\Lambda_0$, and vice versa. Moreover, ${k}$ and $\Lambda_0$ must have \emph{opposite signs}, and \emph{none of them can be zero}.

Since ${\Spq \neq 0}$, the field equations (\ref{GB_pq_s}) have to be satisfied for every spatial component $p$ and~$q$. This implies at most quadratic dependence of ${g_{uu}}$ on ${r}$, similarly as in (\ref{GB_g_uu}), but \emph{without the constraint} (\ref{GB_b_Explicit}) on $b$. Moreover, the value of the transverse-space tensors in (\ref{GB_pq_s}) has to be equal for every ${pq}$-component, i.e., by integrating the equations for all choices of $p,q$ we must obtain \emph{the same} unique  ${g_{uu}}$. We can also substitute the explicit expression for the constant Ricci scalar ${\sR = 4 \Lambda_0}$, together with the generic quadratic form (\ref{GB_g_uu}) of $g_{uu}$ into the ${up}$- and ${uu}$-component of the field equations, see (\ref{GB_up_s}) and (\ref{GB_uu_s}), respectively. In such a peculiar case, these equations remain very similar to those presented in Section~\ref{SubS:neq0}.


\subsection{Case ${\Spq =0}$ implying ${Q = 0}$}
\label{SubS:00}

As in the previous case, the condition ${Q = 0}$ implies the constraints (\ref{GB_Ein_R}) and (\ref{Rje4Lambda}), i.e., (\ref{GBParam}). Moreover, the additional condition ${\Spq =0}$, where $\Spq$ is defined as (\ref{Spq}), puts a  further strong constraint on the transverse-space geometry, namely
\begin{equation}
\sR_{pq} = \frac{1}{4k}\,g_{pq} + \frac{1}{\,2}\sR\,g_{pq} \,. \label{Spq=0}
\end{equation}
Because the spatial Ricci scalar is simply ${\sR=4\Lambda_0 }$, using the coupling (\ref{GBPar}) we obtain
\begin{align}
\sR_{pq} =  \frac{4\Lambda_0}{D-2}\,g_{pq}
    \equiv -\frac{1}{2k(D-4)}\,g_{pq} \,. \label{GB_Ein_R_pq}
\end{align}
We thus have proved that in such a case the ${(D-2)}$-dimensional \emph{transverse-space has to be the Einstein space}. As we have already mentioned, this subclass of vacuum solutions is \emph{not allowed} in Einstein's gravity theory corresponding to ${k=0}$.

Now, we may proceed with the discussion of the remaining field equations. By putting ${\Spq = 0}$ and substituting  (\ref{GB_Ein_R}) and (\ref{GB_Ein_R_pq}) into the general  $pq$-equation (\ref{GB_pq_s}), we obtain the following constraint for the contraction of the transverse-space Riemann tensor,
\begin{align}
\sR_{pklm}\sR_q{}^{klm}
  &=\frac{2}{\big[2k(D-4)\big]^2}\,g_{pq}
   \equiv\frac{32\,\Lambda_0^2}{(D-2)^2}\,g_{pq}  \nonumber\\
  &\equiv\frac{8\Lambda_0}{D-2}\sR_{pq} \equiv -\frac{1}{k(D-4)}\sR_{pq}   \,. \label{GB_Ein_R_pqmn}
\end{align}

With (\ref{Rje4Lambda}), (\ref{GB_Ein_R_pq}), (\ref{GB_Ein_R_pqmn}) and (\ref{GBPar}), the ${ru}$-equation (\ref{GB_ru}) is now identically satisfied.

For ${\Spq = 0}$ and (\ref{GB_Ein_R_pq}), the $up$-equation (\ref{GB_up_s}) simplifies to
\begin{align}
\bigg(-\frac{8\,\Lambda_0}{D-2}\,\delta_p^m\,g^{kl}+\sR_p{}^{kml}\bigg)\,g_{k[m,u||l]}=0 \,. \label{GB_up_sSPEC}
\end{align}
The expression in the round brackets cannot be zero because otherwise the resulting Ricci tensor would be incompatible with~(\ref{GB_Ein_R_pq}).

Finally, the $uu$-component (\ref{GB_uu_s}) for ${\Spq = 0}$ reduces to
\begin{align}
\big(g^{ko}g^{ls}-2\,g^{kl}g^{os}\big)g^{pq}\,g_{k[p,u||l]}\,g_{o[q,u||s]}=0 \,,
\label{GB_uu_sSPEC}
\end{align}
which represents a further constraint for the spatial part $g_{pq}$ of the metric and its $u$-dependence.

We conclude that for this specific subclass of Einstein--Gauss--Bonnet Kundt spacetimes, the parameters of the theory are constrained by the condition (\ref{GBPar}). The transverse space must be an \emph{Einstein space} of the form (\ref{GB_Ein_R_pq}), implying (\ref{Rje4Lambda}). The spatial metric is further constrained by (\ref{GB_Ein_R_pqmn}), (\ref{GB_up_sSPEC}) and (\ref{GB_uu_sSPEC}).

On the other hand, the metric component ${g_{uu}(r,u,x)}$ \emph{remains a fully arbitrary function of all spacetime variables}, i.e., there is no constraint imposed by the field equations.

It can also be immediately observed that the complicated field equations (\ref{GB_up_sSPEC}) and (\ref{GB_uu_sSPEC}) are \emph{trivially satisfied} when the spatial metric $g_{pq}$ is \emph{independent of the retarded time coordinate}~$u$. In such a case, the vacuum solutions to Einstein--Gauss--Bonnet gravity theory with non-zero parameters
\begin{align}
8k\Lambda_0 = -\frac{D-2}{D-4} \ne 0 \,,
\label{SPECsoluionPARAM}
\end{align}
see (\ref{GBParam}), are
\begin{equation}
\dd s^2 = g_{pq}(x)\, \dd x^p\,\dd x^q -2\,\dd u\,\dd r+g_{uu}(r,u,x)\, \dd u^2 \,, \label{Kundt_g(x)}
\end{equation}
where the spatial metric $g_{pq}(x)$ is \emph{any} Einstein space satisfying
\begin{align}
\sR_{pq} =  \frac{4\Lambda_0}{D-2}\,g_{pq}
    \quad\Rightarrow\quad \sR=4\Lambda_0\,, \label{GB_Ein_R_pqg(x)}
\end{align}
together with the specific curvature constraint
\begin{align}
\sR_{pklm}\sR_q{}^{klm}
  & = \frac{8\Lambda_0}{D-2}\sR_{pq}  \equiv  \frac{32\,\Lambda_0^2}{(D-2)^2}\,g_{pq} \,. \label{GB_Ein_R_pqmn(x)}
\end{align}
Notice that the corresponding transverse-space \emph{Kretschmann scalar invariant} is
\begin{align}
\sR_{pklm}\sR^{pklm} & =  \frac{32\,\Lambda_0^2}{D-2} \,. \label{Kretschman-qmn(x)}
\end{align}
It is \emph{everywhere the same and finite}, uniquely determined just by the value of the cosmological constant ${\Lambda_0\ne0 }$. This indicate that the solutions are (in this sense) uniform and non-singular.


\section{Algebraic structure of the Weyl and Ricci tensors}
\label{Sec:AlgStr}

In this section we analyze the algebraic structure of the Weyl and Ricci tensors of the three classes of spacetimes introduced in subsections~\ref{SubS:neq0},~\ref{SubS:0neq0} and~\ref{SubS:00}. We apply the classification scheme of tensors in terms of their boost-weight irreducible components with respect to a suitable null frame \cite{OrtaggioPravdaPravdova:2013, PodolskySvarc:2013a}. Such a natural null frame ${\{\boldk,\,\boldl,\,\boldm_i\}}$ satisfying the normalization conditions ${\boldk\cdot\boldl=-1}$ and ${\boldm_i\cdot\boldm_j=\delta_{ij}}$ (which means ${g_{pq}\,m^p_im^q_j=\delta_{ij}}$), adapted to the Kundt geometry (\ref{Kundt_non_gyr}), is
\begin{equation}
\boldk=\mathbf{\partial}_r \,, \ \qquad \boldl=\tfrac{1}{2}\,g_{uu}\mathbf{\partial}_r+\mathbf{\partial}_u \,, \ \qquad \boldm_i=m_i^p\,\mathbf{\partial}_p \,. \label{Nat_null_frame}
\end{equation}
Following the \emph{Weyl tensor decomposition} \cite{OrtaggioPravdaPravdova:2013}, together with explicit results in the case of Kundt geometries \cite{PodolskySvarc:2013a,PodolskySvarc:2015a}, we introduce the frame components with respect to the generic null frame ${\{\boldk,\,\boldl,\,\boldm_i\}}$ by
\begin{equation}
\begin{aligned}
\Psi_{0^{ij}} &= C_{abcd}\; k^a\, m_i^b\, k^c\, m_j^d \,,&&  \\
\Psi_{1^{ijk}}&= C_{abcd}\; k^a\, m_i^b\, m_j^c\, m_k^d \,,&
\qquad \Psi_{1T^{i}} &= C_{abcd}\; k^a\, l^b\, k^c\, m_i^d \, \\
\Psi_{2^{ijkl}} &= C_{abcd}\; m_i^a\, m_j^b\, m_k^c\, m_l^d \,,&
\qquad \Psi_{2S} &= C_{abcd}\; k^a\, l^b\, l^c\, k^d \,, \\
\Psi_{2^{ij}} &= C_{abcd}\; k^a\, l^b\, m_i^c\, m_j^d \,,&
\qquad \Psi_{2T^{ij}} &= C_{abcd}\; k^a\, m_i^b\, l^c\, m_j^d \,,  \\
\Psi_{3^{ijk}} &= C_{abcd}\; l^a\, m_i^b\, m_j^c\, m_k^d \,,&
\qquad \Psi_{3T^{i}} &= C_{abcd}\; l^a\, k^b\, l^c\, m_i^d \,, \\
\Psi_{4^{ij}} &= C_{abcd}\; l^a\, m_i^b\, l^c\, m_j^d \,. &&
\end{aligned}
\label{Def_Psi}
\end{equation}
These scalars are sorted by their boost weights. Moreover, their irreducible components (which identify specific algebraic subtypes) are
\begin{equation}
\begin{aligned}
\tilde{\Psi}_{1^{ijk}}&= \Psi_{1^{ijk}}-\frac{2}{D-3}\delta_{i[j}\Psi_{1T^{k]}} \,, \\
\tilde{\Psi}_{2T^{(ij)}}&= \Psi_{2T^{(ij)}}-\frac{1}{D-2}\delta_{ij}\Psi_{2S} \,, \\
\tilde{\Psi}_{2^{ijkl}} &= \Psi_{2^{ijkl}}-\frac{2}{D-4}\big(\delta_{ik}\tilde{\Psi}_{2T^{(jl)}}+\delta_{jl}\tilde{\Psi}_{2T^{(ik)}}-\delta_{il}\tilde{\Psi}_{2T^{(jk)}}-\delta_{jk}\tilde{\Psi}_{2T^{(il)}}\big)  -\frac{4\,\delta_{i[k}\delta_{l]j}}{(D-2)(D-3)}\Psi_{2S} \,, \\
\tilde{\Psi}_{3^{ijk}}&=\Psi_{3^{ijk}}-\frac{2}{D-3}\delta_{i[j}\Psi_{3T^{k]}} \,.
\end{aligned}
\label{Def_Irred}
\end{equation}
Evaluating these  quantities for the non-gyratonic Kundt metric (\ref{Kundt_non_gyr}) in the natural null frame (\ref{Nat_null_frame}) we find that the $+2$ and $+1$ boost-weight components $\Psi_{0}$ and $\Psi_{1}$ are identically zero. These geometries are thus \emph{at least of algebraic type}~II, with ${\boldk=\mathbf{\partial}_r}$ being a \emph{double degenerate Weyl-aligned null direction} (WAND). In fact, since also ${\Psi_{2^{ij}}=0}$, see \cite{PodolskySvarc:2013a,PodolskySvarc:2015a}, it is of the \emph{algebraic subtype}~II(d). The remaining Weyl scalars are in general non-trivial and take the form
\begin{align}
\Psi_{2S} & = \frac{D-3}{D-1}\left[\,\tfrac{1}{2}\,g_{uu,rr}+\frac{1}{(D-2)(D-3)}\sR\right] , \label{Psi2S} \\
\hspace{-2.0mm}
\tilde{\Psi}_{2T^{(ij)}} & = m_i^pm_j^q\,\frac{1}{D-2}\left[\sR_{pq}-\frac{1}{D-2}\,g_{pq}\sR \right] , \label{Psi2T} \\
\tilde{\Psi}_{2^{ijkl}} & = m_i^m m_j^p m_k^n m_l^q\,\,^{S}C_{mpnq} \,, \label{Psi2C} \\
\Psi_{3T^{i}} & = m_i^p\,\frac{D-3}{D-2}\left[-\tfrac{1}{2}\,g_{uu,rp}+\frac{1}{D-3}\,g^{mn}g_{m[n,u||p]}\right] , \label{Psi3T} \\
\tilde{\Psi}_{3^{ijk}} & = m_i^pm_j^mm_k^q\left[\,g_{p[m,u||q]}-\frac{1}{D-3}\,g^{os}\left(g_{pm}g_{o[s,u||q]}-g_{pq}g_{o[s,u||m]}\right)\right] , \label{Psi3} \\
\Psi_{4^{ij}} & = m_i^pm_j^q\left[-\tfrac{1}{2}g_{uu||pq}-\tfrac{1}{2}g_{pq,uu}+\tfrac{1}{4}g^{os}g_{op,u}g_{sq,u}+\tfrac{1}{4}g_{pq,u}g_{uu,r}\right. \nonumber \\
& \hspace{20.0mm} \left. -\frac{g_{pq}}{D-2}\,g^{mn}\left(-\tfrac{1}{2}g_{uu||mn}-\tfrac{1}{2}g_{mn,uu}+\tfrac{1}{4}g^{os}g_{om,u}g_{sn,u}+\tfrac{1}{4}g_{mn,u}g_{uu,r}\right)\right]. \label{Psi4ij}
\end{align}
These results apply to \emph{any Kundt geometry}. For solutions to specific gravity theory, the Weyl scalars  have to be further expressed using the corresponding field equation constraints. In the \emph{Einstein--Gauss--Bonnet gravity} we thus obtain:
\begin{itemize}
\item In the generic case ${Q \neq 0}$ of subsection~\ref{SubS:neq0}, the main modification arises from the explicit form (\ref{GB_g_uu}) of the $g_{uu}$ metric function, quadratic in $r$-coordinate. In particular, ${\tfrac{1}{2}\,g_{uu,rr}=b}$ given by (\ref{GB_b_Explicit}). However, the algebraic type~II(d) of the Kundt solution remains in general unchanged.

\item The case ${Q = 0}$ with ${\Spq \neq 0}$, discussed in subsection~\ref{SubS:0neq0}, is even less restrictive than the case ${Q \neq 0}$. Its algebraic type remains II(d). It specializes to II(ad) if, and only if, ${\Psi_{2S}=0}$. Due to (\ref{Psi2S}), (\ref{Rje4Lambda}), (\ref{GB_Ein_R}) this occurs when
\begin{equation}
b = -\frac{\sR}{(D-2)(D-3)} = -\frac{4\Lambda_0}{(D-2)(D-3)} = \frac{1}{2k(D-3)(D-4)} \,. \label{typeII(ad)}
\end{equation}

\item In the class ${\Spq = 0}$ implying ${Q =0}$, see subsection~\ref{SubS:00}, both the highest admitted and the lowest boost-weight components $\Psi_{2}$ and $\Psi_{4}$ contain an \emph{arbitrary} metric function $g_{uu}$. They are thus in general non-vanishing, so that the algebraic (sub)type of Kundt spacetimes (\ref{Kundt_non_gyr}) has to be II(d), or of a more special subtype. Indeed, due to the conditions (\ref{GB_Ein_R}) and (\ref{GB_Ein_R_pq}) specifying the transverse Einstein space we get ${\tilde{\Psi}_{2T^{(ij)}}=0}$, and the Weyl type specializes to II(bd).
\end{itemize}
The explicit form of the scalars (\ref{Psi2S})--(\ref{Psi4ij}) can be employed to discuss the specific algebraically special subclasses within the Kundt solutions (\ref{Kundt_non_gyr}) in the Einstein--Gauss--Bonnet gravity. For example, the scalars (\ref{Psi2S})--(\ref{Psi2C}) imply that the geometry becomes of the Weyl-type~III or more special if, and only if, the transverse space is conformally flat (${\tilde{\Psi}_{2^{ijkl}}=0}$) Einstein space (${\tilde{\Psi}_{2T^{(ij)}}=0}$) and ${g_{uu}}$ function is at most quadratic in $r$, with the coefficient of ${r^2}$ proportional to the spatial curvature ${\sR}$ (to obtain ${\Psi_{2S} =0}$).

We can also define the \emph{traceless Ricci tensor} ${\cR_{ab}\equiv R_{ab}-\frac{1}{D}\,Rg_{ab}}$. Its frame components ${\Phi_{AB}}$ with respect to the null frame ${\{\boldk,\,\boldl,\,\boldm_i\}}$ given by (\ref{Nat_null_frame}), evaluated using the explicit coordinate components (\ref{Ricci_all}) for the Kundt metric~(\ref{Kundt_non_gyr}), are
\begin{align}
\Phi_{00}=&\frac{1}{2}\,\cR_{ab}\;k^a\,k^b=0 \,, \label{Phi00nat}\\
\Phi_{01^i}=&\frac{1}{\sqrt{2}}\;\cR_{ab}\,k^a\,m_i^b=0 \,, \\
\Phi_{11}=&\cR_{ab}\;k^a\,l^b=-\frac{1}{2}g_{uu,rr}+\frac{1}{D}\Big(\sR+g_{uu,rr}\Big) , \label{Phi11nat}\\
\Phi_{02^{ij}}=&\cR_{ab}\;m_i^a\,m_j^b=m_i^p\,m_j^q\,\sR_{pq}-\frac{1}{D}\Big(\sR+g_{uu,rr}\Big)\delta_{ij} \,, \label{Phi02ijnat} \\
\Phi_{12^i}=&\frac{1}{\sqrt{2}}\,\cR_{ab}\;l^a\,m_i^b=\frac{1}{\sqrt{2}}\,m_i^p\Big(-\frac{1}{2}\,g_{uu,rp}+g^{mn}g_{m[p,u||n]}\Big)  \,, \\
\Phi_{22}=&\frac{1}{2}\,\cR_{ab}\;l^a\,l^b=\frac{1}{8}\,g^{mn}g_{mn,u}g_{uu,r}-\frac{1}{4}\,g^{mn}g_{mn,uu} -\frac{1}{4}\,g^{mn}g_{uu||mn}+\frac{1}{8}\,g^{mn}g^{pq}g_{pm,u}g_{qn,u}  \label{Phi22nat} \,.
\end{align}
Because ${\Phi_{00}=0=\Phi_{01^i}}$, the metric ansatz (\ref{Kundt_non_gyr}) always leads to the \emph{algebraically special Ricci tensor}.

To analyze the genuine Gauss--Bonnet contribution, the above expressions have to be further modified using the constraints implied by the field equations~(\ref{RicciFieldEqs}). To this end, it is convenient to rewrite the non-trivial Ricci components using the field equations with the term ${H_{ab}}$ defined in (\ref{GB_FE_contrib}). Its decomposition into the trace $H$, see (\ref{TraceFieldEqs}), and the traceless part ${\cH_{ab}\equiv H_{ab}-\frac{1}{D}Hg_{ab}}$ leads to the relation ${\cR_{ab}=-2k\,\cH_{ab}}$, so that
\begin{align}
\Phi_{11}=&-2k\,\cH_{ru} \,, \label{Phi11natFE} \\
\Phi_{02^{ij}}=&-2k\,m_i^pm_j^q\,\cH_{pq} \,, \label{Phi02ijnatFE}\\
\Phi_{12^i}=&-\sqrt{2}k\,m_i^p\, \cH_{up} \,, \\
\Phi_{22}=&-k\left(g_{uu}\cH_{ru}+\cH_{uu}\right) \label{Phi22natFE} \,.
\end{align}

To obtain an \emph{algebraically more special Ricci tensor}, both its zero-boost-weight components given by (\ref{Phi11nat}) and (\ref{Phi02ijnat}) have to vanish, that is
\begin{equation}
\Phi_{11}=0 \,,\qquad\hbox{and}\qquad \Phi_{02^{ij}}=0 \,.\label{specialRicci}
\end{equation}
We study the solutions in Einstein--Gauss--Bonnet gravity, and therefore the corresponding conditions (\ref{Phi11natFE}), (\ref{Phi02ijnatFE}) implied by the field equations must be zero. The condition (\ref{Phi11nat}) implies ${g_{uu,rr}=2\sR/(D-2)}$, that is
\begin{equation}
  g_{uu}=\frac{\sR}{D-2}\,r^2+c\,r+d\,, \quad\qquad \sR_{pq}=\frac{\sR}{D-2}\,g_{pq} \,, \label{Ricci_spec_1}
\end{equation}
while its combination with the second condition (\ref{Phi02ijnatFE}) gives
\begin{equation}
\sR^2_{klmn}=2\frac{\sR^2}{D-2} \,, \quad\qquad \sR_{pklm}{\sR_{q}}^{klm}=2\frac{\sR^2}{(D-2)^2}\,g_{pq} \,, \label{Ricci_spec_2}
\end{equation}
where we have employed the explicit expressions for ${H_{ab}}$ and its trace $H$, given in (\ref{Hru})--(\ref{HExpl}). Obviously, these constraints also \emph{specialize the Weyl tensor} to type II(bd) since ${\tilde{\Psi}_{2T^{(ij)}}=0}$, see (\ref{Psi2T}).


\section{Geodesic deviation in the Einstein--Gauss--Bonnet theory}
\label{Sec:GeodDev}

The specific tidal deformations caused by inhomogeneities of the gravitational field can be naturaly observed via their influence on \emph{freely falling nearby test particles}, such as the test masses of the LISA detector. Geometrically, these effects are encoded in the spacetime curvature ${R^a_{\ bcd}}$, and described by the equation of geodesic deviation,
\begin{equation}\label{EqGeoDev}
\frac{{\rm{D}}^2 Z^a}{\dd\tau^2}=R^a_{\ bcd}\,u^b u^c Z^d \,,
\end{equation}
where $u^b$ are components of the reference observer velocity, which moves along a time-like geodesic ${\gamma(\tau)}$ with~${\tau}$ being its proper time, and $Z^a$ are components of the vector connecting this observer with another one moving nearby. To obtain an invariant description \cite{Pirani:1957, BondiPiraniRobinson:1959, Szekeres:1965, BicakPodolsky:1999b, PodolskySvarc:2012} of such tidal deformations, we employ an orthonormal frame ${\{\bolde_{(0)},\,\bolde_{(1)},\,\bolde_{(i)}\}}$ associated with the fiducial test observer, i.e., ${\bolde_a \cdot \bolde_b=\eta_{ab}}$, where we assume ${\bolde_{(0)}\equiv\boldu=\dot{r}\,\mathbf{\partial}_{r}+\dot{u}\,\mathbf{\partial}_{u}+\dot{x}^p\mathbf{\partial}_{p}\,}$. The projection of equation (\ref{EqGeoDev}) onto such a frame can be written as ${\ddot Z^{(a)}= R^{(a)}_{\quad(0)(0)(b)}\,Z^{(b)}}$ with ${\ddot Z^{(a)} \equiv e^{(a)}_b\,\frac{{\rm{D}}^2Z^b}{\dd\, \tau^2}}$ and ${Z^{(b)}\equiv e^{(b)}_aZ^a}$, where ${a,b=0,1,\ldots, D-1}$. This immediately gives ${\ddot Z^{(0)}=0}$ and, without loss of generality, we can set ${Z^{(0)}=0}$ corresponding to the test observers always located at the same spacelike hypersurfaces synchronized by their proper time $\tau$. Subsequently, using a standard decomposition of the Riemann tensor \cite{Wald}, the invariant form of the equation of geodesic deviation becomes
\begin{equation}\label{InvGeoDev}
\ddot Z^{(\rm{i})}= \left[C_{(\rm{i})(0)(0)(\rm{j})}+\frac{1}{D-2}\left(R_{(\rm{i})(\rm{j})}-\delta_{\rm{i}\rm{j}}\,R_{(0)(0)}\right) -\frac{R\,\delta_{\rm{i}\rm{j}} }{(D-1)(D-2)}\right]Z^{(\rm{j})} \,,
\end{equation}
where ${{\rm{i}},{\rm{j}}=1,2,\ldots, D-1}$. To analyze particular contributions to the total deformation of a test congruence, we define the \emph{null interpretation frame} as
\begin{equation}
\boldk^\intf={\textstyle\frac{1}{\sqrt{2}}}(\boldu+\bolde_{(1)}) \,, \qquad \boldl^\intf={\textstyle\frac{1}{\sqrt{2}}}(\boldu-\bolde_{(1)}) \,, \qquad
\boldm_{i}^\intf=\bolde_{(i)} \,. \label{IntNullFrameDef}
\end{equation}
Then the Weyl tensor projections can be expressed in term of the scalars (\ref{Def_Psi}) as
\begin{align}
C_{(1)(0)(0)(1)} =& \,\Psi_{2S}^\intf \,, \nonumber \\
C_{(1)(0)(0)(j)} =& \,{\textstyle\frac{1}{\sqrt{2}}}(\,\Psi_{1T^j}^\intf-\Psi_{3T^j}^\intf) \,, \nonumber \\
C_{(i)(0)(0)(1)} =& \,{\textstyle\frac{1}{\sqrt{2}}}(\,\Psi_{1T^i}^\intf-\Psi_{3T^i}^\intf) \,, \label{OrthoNullWeylRelations} \\
C_{(i)(0)(0)(j)} =& \,-{\textstyle\frac{1}{2}}(\,\Psi_{0^{ij}}^\intf+\Psi_{4^{ij}}^\intf) - \Psi_{2T^{(ij)}}^\intf \,, \nonumber
\end{align}
and for the relevant Ricci tensor components using the definitions (\ref{Phi00nat})--(\ref{Phi22nat}) we obtain
\begin{align}
R_{(0)(0)} =& \,\Phi_{00}^\intf+\Phi_{22}^\intf+\Phi_{11}^\intf-\frac{R}{D} \,, \nonumber \\
R_{(1)(1)} =& \,\Phi_{00}^\intf+\Phi_{22}^\intf-\Phi_{11}^\intf+\frac{R}{D} \,, \nonumber \\
R_{(1)(j)} =& \,\Phi_{01^j}^\intf-\Phi_{12^j}^\intf \,, \label{OrthoNullRicciRelations} \\
R_{(i)(j)} =& \,\Phi_{02^{ij}}^\intf+\frac{R}{D}\,\delta_{ij} \,, \nonumber
\end{align}
with ${\,i,j=2,\ldots,D-1\,}$ labelling $D-2$ spatial directions orthogonal to the privileged longitudinal direction~${\bolde_{(1)}}$.

By combining the definition (\ref{IntNullFrameDef}) with the orthonormality condition ${\bolde_a \cdot \bolde_b=\eta_{ab}}$ we obtain the explicit form of the null interpretation frame, namely
\begin{align}
\boldk^\intf &= \frac{1}{\sqrt{2}\,\dot{u}}\,\mathbf{\partial}_r \, , \nonumber \\
\boldl^\intf &= \Big(\sqrt{2}\,\dot{r}-\frac{1}{\sqrt{2}\,\dot{u}}\Big)\mathbf{\partial}_r+\sqrt{2}\,\dot{u}\,\mathbf{\partial}_u+\sqrt{2}\,\dot{x}^p\mathbf{\partial}_p \, , \label{Kundt interpretation frame} \\
\boldm_i^\intf &= \frac{1}{\dot{u}}\,g_{pq}\,m_{i}^{p}\,\dot{x}^q\,{\partial}_r+m_i^p\,\mathbf{\partial}_p \, . \nonumber
\end{align}
Using the Lorentz transformation we may relate this interpretation frame (adapted to a generic time-like observer) with the natural null frame (\ref{Nat_null_frame}) corresponding to the choice of a specific static observer with ${\,\sqrt{2}\,\dot{u}=1,\, \dot{x}^p=0\,}$ (and ${\,\sqrt{2}\,\dot{r}-1=\frac{1}{2}g_{uu}}$ due to ${\boldu \cdot \boldu=-1}$), see \cite{PodolskySvarc:2012, PodolskySvarc:2013b} for more details. In particular, it is a combination of a boost followed by a null rotation with fixed $\boldk $,
\begin{align}
\boldk^\intf &= B\boldk \, , \nonumber \\
\boldl^\intf &= B^{-1}\boldl  + \sqrt2\,L^i \boldm_i  + |L|^2 B\boldk \, , \label{Lorentztransf}\\
\boldm_i^\intf &= \boldm_i  + \sqrt2\,L_i \,B\boldk \, , \nonumber
\end{align}
where ${|L|^2\equiv \delta^{ij}L_iL_j}$ and
\begin{equation}
B=\frac{1}{\sqrt2\,\dot{u}}\,, \qquad L_i=g_{pq}\,m_{i}^{p}\,\dot{x}^q\,.
\label{specialchoiceofframe}
\end{equation}
Using the Lorentz transformation (\ref{Lorentztransf}) and definition (\ref{Def_Psi}), we can evaluate the Weyl scalars in the decomposition (\ref{OrthoNullWeylRelations})  with respect to the null interpretation frame (\ref{Kundt interpretation frame}) in terms of the scalars (\ref{Psi2S})--(\ref{Psi4ij}) with (\ref{Def_Irred}) expressed in the natural null frame (\ref{Nat_null_frame}) adapted to the algebraic structure of the spacetime. It turns out that
\begin{align}
{\Psi}_{0^{ij}}^\intf &= 0 \,, \qquad {\Psi}_{1T^{i}}^\intf = 0 \,, \qquad {\Psi}_{2S}^\intf = \Psi_{2S} \,, \qquad {\Psi}_{2T^{ij}}^\intf = \Psi_{2T^{ij}}  \,, \nonumber \\
{\Psi}_{3T^{i}}^\intf &= B^{-1}\Psi_{3T^{i}}-\sqrt{2}\left(\Psi_{2T^{ki}}L^k+\Psi_{2S}L_i\right) \,, \nonumber\\
{\Psi}_{4^{ij}}^\intf &= B^{-2}\Psi_{4^{ij}} +2\sqrt{2}\,B^{-1}\!\left(\Psi_{3T^{(i}}L_{j^{)}}-\Psi_{3^{(ij)k}}L^k\right) \nonumber \\
& \quad + 2\Psi_{2^{ikjl}}L^kL^l -4\Psi_{2T^{k(i}}L_{j^{)}}L^k +2\Psi_{2T^{(ij)}}|L|^2  - 2\Psi_{2S}L_iL_j \,. \label{Psi_inter_nat}
\end{align}
Employing the same procedure, the Ricci tensor frame components, entering the projection (\ref{OrthoNullRicciRelations}), expressed using those with respect to the natural frame (\ref{Phi00nat})--(\ref{Phi22nat}) become
\begin{align}
\Phi_{00}^\intf &=0 \,, \qquad \Phi_{01^j}^\intf =0 \,, \qquad \Phi_{11}^\intf =\Phi_{11} \,, \qquad \Phi_{02^{ij}}^\intf =\Phi_{02^{ij}} \,, \nonumber \\
\Phi_{12^j}^\intf &= B^{-1}\Phi_{12^j}+\Phi_{11}L_j+\Phi_{02^{ij}}L^i \,, \label{Ricci_inter_nat} \\
\Phi_{22}^\intf &= B^{-2}\Phi_{22}+2B^{-1}\Phi_{12^j}L^j+\Phi_{11}|L|^2+\Phi_{02^{ij}}L^iL^j \,. \nonumber
\end{align}
In general, all these scalars have to be evaluated as functions of proper time $\tau$ along the fiducial observer geodesic ${\gamma(\tau)}$. However, for the physical analysis of the spacetime geometry one can use them in a local sense where their values at any given event correspond to the actual accelerations of test observers.

Finally, we can now explicitly write down the \emph{invariant form of geodesic deviation equations} for a generic time-like observer freely falling in the non-gyratonic Kundt geometries (\ref{Kundt_non_gyr}),
\begin{align}
\ddot Z^{(1)}=& \frac{R}{D(D-1)}\,Z^{(1)}+\Psi_{2S}^\intf\,Z^{(1)}-\frac{1}{\sqrt{2}}\,\Psi_{3T^j}^\intf\,Z^{(j)} \nonumber \\
& \hspace{10mm} -\frac{1}{D-2}\left(2\Phi_{11}^\intf\,Z^{(1)}+\Phi_{12^j}^\intf\,Z^{(j)}\right), \label{InvGeoDevExplicit_1} \\
\ddot Z^{(i)}=& \frac{R}{D(D-1)}\,Z^{(i)}-\Psi_{2T^{(ij)}}^\intf\,Z^{(j)}-\frac{1}{\sqrt{2}}\,\Psi_{3T^i}^\intf\,Z^{(1)}-\frac{1}{2}\Psi_{4^{ij}}^\intf\,Z^{(j)} \nonumber \\
& \hspace{10mm} -\frac{1}{D-2}\left(-\Phi_{02^{ij}}^\intf\,Z^{(j)}+\Phi_{12^i}^\intf\,Z^{(1)}+\left(\Phi_{22}^\intf+\Phi_{11}^\intf\right)Z^{(i)}\right), \label{InvGeoDevExplicit_i}
\end{align}
where ${i,j=2,\ldots,D-1}$. The scalar curvature can be expressed using (\ref{TraceFieldEqs}), and the remaining Weyl and Ricci scalars are given by (\ref{Psi_inter_nat}) and (\ref{Ricci_inter_nat}) with (\ref{Psi2S})--(\ref{Psi4ij}) and (\ref{Phi11natFE})--(\ref{Phi22natFE}), respectively, together with (\ref{specialchoiceofframe}).

We observe that the \emph{Gauss--Bonnet terms} encoded via the vacuum field equations (\ref{RicciFieldEqs}) in the Ricci tensor components $\Phi^\intf_{AB}$ cause specific relative accelerations of free test observers. This is an \emph{additional contribution to the Weyl tensor components} ${\Psi^\intf_{A}}$ representing the only relevant effects of the gravitational field in the vacuum Einstein theory.

\newpage

In particular, the term proportional to ${R \equiv \sR+g_{uu,rr}}$, see (\ref{Ricci_scalar}), determines the \emph{isotropic	influence} of the cosmological constant $\Lambda_0$ combined with the direct contribution of the Gauss--Bonnet term $L_{GB}$ via the trace~$H$ as ${R = \frac{2}{D-2}(D\Lambda_0+2kH)}$, see (\ref{TraceFieldEqs}). In addition, there is the \emph{Newtonian tidal effect}, caused by the ${\Psi_{2S}^\intf}$ and ${\Psi_{2T^{(ij)}}^\intf}$ Weyl components, which specifically deform the test body in all spatial directions due to the constraint ${\Psi_{2S}^\intf=\delta^{ij}\,\Psi_{2T^{(ij)}}^\intf}$. These directions are also \emph{similarly influenced} by  ${2\Phi_{11}^\intf}$ acting in the longitudinal ${\bolde_{(1)}}$ direction, and ${\Phi_{02^{ij}}^\intf}$ affecting the remaining ${(D-2)}$ transverse directions ${\bolde_{(i)}}$. As in the case of the Newtonian effect, these terms satisfy the constraint ${2\Phi_{11}^\intf=\delta^{ij}\,\Phi_{02^{ij}}^\intf}$. Moreover, the \emph{longitudinal} deformations corresponding to ${\Psi_{3T^j}^\intf}$ are similar to the effect of ${\Phi_{12^i}}$. The \emph{purely transverse} deformations classically related to \emph{gravitational waves} are encoded in the Weyl ${\Psi_{4^{ij}}^\intf}$ scalars which are traceless, ${\delta^{ij}\,\Psi_{4^{ij}}^\intf=0}$. Finally, there is also the peculiar combined influence of ${\Phi_{22}^\intf+\Phi_{11}^\intf}$  deforming these transverse directions.

To analyze the specific role of the Gauss--Bonnet theory more explicitly (and suppress the complicating kinematic effect of the observer's motion) we restrict ourselves to the (initially) transversally \emph{static observers}, that is  ${\,\sqrt{2}\,\dot{u}=1,\, \dot{x}^p=0\,}$,  so that ${B=1}$ and ${L_i=0}$. This corresponds to the direct choice of the natural frame (\ref{Nat_null_frame}). By substituting the Ricci tensor contributions $\Phi_{AB}$ from (\ref{Phi11natFE})--(\ref{Phi22natFE}), we obtain
\begin{align}
\ddot Z^{(1)}=& \frac{2\left(\Lambda_0+2kH/D\right)}{(D-1)(D-2)}Z^{(1)}+\Psi_{2S} \,Z^{(1)}-\frac{1}{\sqrt{2}}\,\Psi_{3T^j} \,Z^{(j)} \nonumber \\
&\hspace{10mm}  +\frac{k}{D-2}\left(4\cH_{ru} \,Z^{(1)}+\sqrt{2}m_j^p\, \cH_{up} \,Z^{(j)}\right)  , \label{InvGeoDevExplicit_1_nat} \\
\ddot Z^{(i)}=& \frac{2\left(\Lambda_0+2k H/D\right)}{(D-1)(D-2)}Z^{(i)}-\Psi_{2T^{(ij)}} \,Z^{(j)}-\frac{1}{\sqrt{2}}\,\Psi_{3T^i} \,Z^{(1)}-\frac{1}{2}\Psi_{4^{ij}} \,Z^{(j)} \nonumber \\
&\hspace{10mm}  +\frac{k}{D-2}\left(-2m_i^pm_j^q\,\cH_{pq}\,Z^{(j)}
+\sqrt{2}m_i^p\, \cH_{up} \,Z^{(1)}
+\big[(g_{uu}+2)\cH_{ru}+\cH_{uu}\big]\,Z^{(i)}\right), \label{InvGeoDevExplicit_i_nat}
\end{align}
where the components of the traceless Gauss--Bonnet part ${\cH_{ab}}$ satisfy ${2\cH_{ru}=g^{pq}\,\cH_{pq}}$. Explicit form of the Gauss--Bonnet quantities ${\cH_{ab}\equiv H_{ab}-\frac{1}{D}Hg_{ab}}$ can be simply calculated using ${H_{ab}}$ and the trace $H$, presented in (\ref{Hru})--(\ref{Huu}) and (\ref{HExpl}).

\subsection{Example: Solutions of the Ricci type~III}

A better understanding of the specific terms and their mutual couplings in the above equations can be achieved via study of simplified particular examples. Let us assume here the Kundt spacetimes (\ref{Kundt_non_gyr}) with \emph{$u$-independent transverse space metric} ${g_{pq}}$,  and additional constraints corresponding to the vanishing traceless Ricci tensor ${\cR_{ab}}$ zero-boost-weight components ${\Phi_{11}=0}$ and ${\Phi_{02^{ij}}=0}$ which are presented in equations (\ref{Ricci_spec_1})--(\ref{Ricci_spec_2}). In such a case, the expressions (\ref{InvGeoDevExplicit_1_nat}) and (\ref{InvGeoDevExplicit_i_nat}) for the geodesic deviation reduce to
\begin{align}
\ddot Z^{(1)}=& \frac{2\left(\Lambda_0+2kH/D\right)}{(D-1)(D-2)}\,Z^{(1)}+\Psi_{2S} \,Z^{(1)}-\frac{1}{\sqrt{2}}\,\Psi_{3T^j} \,Z^{(j)} \nonumber \\
&\hspace{32.5mm}  +\frac{k}{D-2}\,\sqrt{2}m_j^p\, \cH_{up} \,Z^{(j)}  \,, \label{InvGeoDevExplicit_1_nat_spec} \\
\ddot Z^{(i)}=& \frac{2\left(\Lambda_0+2k H/D\right)}{(D-1)(D-2)}\,Z^{(i)}-\frac{1}{D-2}\Psi_{2S} \,Z^{(i)}-\frac{1}{\sqrt{2}}\,\Psi_{3T^i} \,Z^{(1)}-\frac{1}{2}\Psi_{4^{ij}} \,Z^{(j)} \nonumber \\
&\hspace{32.5mm}  +\frac{k}{D-2}\left(\sqrt{2}m_i^p\, \cH_{up} \,Z^{(1)}
+\cH_{uu}\,Z^{(i)}\right), \label{InvGeoDevExplicit_i_nat_spec}
\end{align}
with the Weyl tensor components
\begin{align}
\Psi_{2S} & = \frac{1}{D-1}\,\sR\, , \label{Psi2S_spec} \\
\hspace{-2.0mm}
\Psi_{3T^{i}} & = -\frac{1}{2}\frac{D-3}{D-2}\,m_i^p\,g_{uu,rp}\,, \label{Psi3T_spec} \\
\Psi_{4^{ij}} & = -\frac{1}{2}m_i^pm_j^q\left(g_{uu||pq} -\frac{g_{pq}}{D-2}\,g^{mn}g_{uu||mn}\right). \label{Psi4ij_spec}
\end{align}
The traceless Ricci tensor contributions are
\begin{align}
\cH_{up} =& -\frac{1}{2}\frac{D-4}{D-2}\,\sR\, g_{uu,rp} \,, \qquad \cH_{uu} = -\frac{1}{2}\frac{D-4}{D-2}\,\sR\, g^{pq}\,g_{uu||pq}  \,, \label{cHup_cHuu_spec}
\end{align}
and the trace part is
\begin{equation}
H= -\frac{D(D-4)}{4(D-2)}\,\sR^2\,. \label{HExpl_spec}
\end{equation}
They clearly vanish for ${D=4}$, reducing (\ref{InvGeoDevExplicit_1_nat_spec}), (\ref{InvGeoDevExplicit_i_nat_spec}) to the results known for standard general relativity \cite{BicakPodolsky:1999b, PodolskySvarc:2012, PodolskySvarc:2013b} with the cosmological term $\frac{1}{3}\Lambda_0$. Moreover, the metric functions are constrained by the remaining field equations. Namely, the trace equation (\ref{TraceFieldEqs}) couples the transverse space scalar curvature $\sR$ to the theory constants,
\begin{equation}
\sR=\frac{D-2}{2k (D-4)}\bigg(\pm\sqrt{1+8k \Lambda_0\frac{D-4}{D-2}}-1\bigg) , \label{sR_Ricc_spec}
\end{equation}
and the ${up}$ and ${uu}$ components give the conditions
\begin{align}
g_{uu,rp}\,\sqrt{1+8k \Lambda_0\frac{D-4}{D-2}} =0 \,, \qquad g^{pq}g_{uu||pq}\,\sqrt{1+8k \Lambda_0\frac{D-4}{D-2}}=0 \,,
\label{specialcase}
\end{align}
respectively. In view of this, there are thus \emph{two classes of solutions} corresponding to subsections~\ref{SubS:neq0} and~\ref{SubS:00}:
\begin{itemize}
\item Generic case with $g_{uu}$ given by (\ref{Ricci_spec_1}), implying necessarily ${g_{uu,rp}=c_{,p}=0}$, and ${g^{pq}g_{uu||pq}=0}$. Under the assumption of this example, this does \emph{not cause any non-classical motion of the test particles}, since ${\cH_{up}=0=\cH_{uu}}$. There is only the background isotropic modification via the trace $H$ to the value
\begin{equation}
2\,(\Lambda_0+2kH/D) = \sR\,. \label{H-adjucted}
\end{equation}
The same constraints are obtained also in the Einstein theory (when ${k=0}$).

\item The special class of solutions corresponding to a specific value of ${\Lambda_0}$, namely
\begin{equation}
\Lambda_0=-\frac{1}{8k}\frac{D-2}{D-4} \,, \label{specific-lambda}
\end{equation}
for which the functions $c$ and $d$  in $g_{uu}$ given by (\ref{Ricci_spec_1}) \emph{remain unconstrained} by (\ref{specialcase}). These terms cause the additional longitudinal effect in (\ref{InvGeoDevExplicit_1_nat_spec}) via the ${\cH_{up}}$ component, and a peculiar transverse deformation in (\ref{InvGeoDevExplicit_i_nat_spec}) generated by ${\cH_{uu}}$. These effects  are \emph{not allowed in classic general relativity} without the Gauss--Bonnet contribution.
\end{itemize}


\section{Geometrically special members of the Kundt class}
\label{Sec:Subclasses}

In this section we concentrate on \emph{two interesting and physically important examples} of the non-gyratonic Kundt metrics (\ref{Kundt_non_gyr}). In the first case, we restrict the geometry of transverse space to be of a constant curvature. In the second case, there is no a priori restriction applied to the transverse space, but the metric is assumed to be $r$-independent which corresponds to the famous {\it pp\,}-wave class of gravitational waves.


\subsection{Waves and backgrounds with a constant-curvature transverse space\label{SubS:ConsTrSpace}}

We employ the general results of section~\ref{Sec:FEqs} to investigate those Kundt geometries of the form (\ref{Kundt_non_gyr}) for which the ${(D-2)}$-dimensional transverse Riemannian space with the metric ${g_{pq}(u,x)}$ has a \emph{constant curvature} implying ${\sR=\hbox{const.}}$ (with respect to the spatial coordinates $x^p$). In such a case the Riemann tensor can be written as
\begin{equation}
\sR_{pqmn}=\frac{\sR}{(D-3)(D-2)}\,(g_{pm}g_{qn}-g_{pn}g_{qm}) \,, \label{const_curv_Riem}
\end{equation}
and for its contractions we immediately get the relations
\begin{equation}
\sR_{klmn}^2=2\,\frac{\sR^2}{(D-3)(D-2)}\,, \qquad \sR_{pq}=\frac{\sR}{D-2}\,g_{pq} \,, \qquad \sR_{mn}^2=\frac{\sR^2}{D-2}\,. \label{const_curv_R_contract}
\end{equation}
Also, the ${(D-2)}$-dimensional transverse metric ${g_{pq}=g_{pq}(u,x)}$ can be written in a conformal form
\begin{align}
g_{pq}=P^{-2}\,\delta_{pq}\,,\qquad \text{where} \qquad P=1+\frac{\sR}{4(D-3)(D-2)}\left[\left(x^2\right)^2 + \cdots + \left(x^{D-1}\right)^2\right] .  \label{const_curv_g}
\end{align}

Now, we can proceed to discussion of the Einstein--Gauss--Bonnet field equations. Substituting (\ref{const_curv_R_contract}) into the ${ru}$-component (\ref{GB_ru}), we obtain the constraint
\begin{equation}
k \,\frac{(D-4)(D-5)}{(D-2)(D-3)}\sR^2+\sR-2\Lambda_0 = 0\,. \label{CCS_GB_ru}
\end{equation}
Here, we assume that the theory parameters ${k=\kappa\gamma}$ and $\Lambda_0$ are \emph{generic} (and non-zero). This equation is thus understood as an \emph{algebraic condition for the scalar curvature} $\sR$.  The constant coefficients in (\ref{CCS_GB_ru}) immediately imply that $\sR$ has to be $u$-independent, which together with (\ref{const_curv_g}) gives ${g_{pq}=g_{pq}(x)}$. Solving (\ref{CCS_GB_ru}), we explicitly and uniquely express the transverse Ricci scalar in terms of the theory parameters $\kappa$, $\gamma$, and $\Lambda_0$ as
\begin{equation}
\sR=\frac{(D-2)(D-3)}{2k (D-4)(D-5)}\left(\pm\sqrt{1+8k \Lambda_0\frac{(D-4)(D-5)}{(D-2)(D-3)}} - 1 \right).\label{sR_CC_GB}
\end{equation}
Obviously, there are exceptional cases ${D=5}$  and ${D=4}$ in (\ref{CCS_GB_ru}) for which ${\sR=2\Lambda_0}$, corresponding to the classic Einstein's theory constraint.

There are \emph{two branches of such exact Kundt solutions} in the Einstein--Gauss--Bonnet gravity. The first for the $``+"$ choice in (\ref{sR_CC_GB}) admits the general relativity limit as ${k \to 0}$ leading to ${\sR=2\Lambda_0}$, while the second with the $``-"$ choice in (\ref{sR_CC_GB}) is peculiar.

The crucial quantity ${\Spq }$ given by (\ref{Spq}), which defines three distinct subclasses of section~\ref{Sec:FEqs}, becomes
\begin{equation}
\Spq = -\frac{1}{2}\left(1+2k\, \frac{D-4}{D-2}\sR\right)g_{pq} \,. \label{Spq_CC}
\end{equation}
For a generic $k$, $\Lambda_0$ and $\sR$ given by (\ref{sR_CC_GB}), its trace ${Q \equiv g^{pq}\Spq }$, see (\ref{SpqTRACE}), is non-vanishing. Therefore, the Kundt spacetimes with constant-curvature transverse space belong to the general class discussed in subsection~\ref{SubS:neq0}. The trace of the field equations ${pq}$-component (\ref{GB_tr_s}) thus then implies (\ref{GB_g_uu}), (\ref{GB_b_Explicit}), that is
\begin{align}
g_{uu}=b\,r^2+c(u, x)\,r+d(u, x) \,, \qquad \hbox{with} \qquad b=\frac{4\Lambda_0-\sR}{(D-2)+2k\, (D-4)\sR} \,. \label{GB_c_g}
\end{align}

Moreover, due to the independence of the spatial metric $g_{pq}$ on $u$-coordinate, the equation (\ref{GB_up_s}) simplifies considerably to
\begin{align}
\left(1+2k \, \frac{D-4}{D-2}\sR\right)g_{uu,rm} = 0 \,,
\end{align}
which, using (\ref{sR_CC_GB}) and (\ref{GB_c_g}), leads to the simple constraint ${c=c(u)}$. This is consistent with equations (\ref{GB_up_1}) and (\ref{GB_up_r0}). Finally, from the ${uu}$-component (\ref{GB_uu_s}) we obtain the condition
\begin{align}
\left(1+2k \,\frac{D-4}{D-2}\sR\right)g^{pq} g_{uu||pq}=0 \,,
\end{align}
with only non-trivial $r$-independent part implying
\begin{align}
 \triangle d \equiv  g^{pq} d_{||pq} =0 \,, \label{laplace}
\end{align}
see (\ref{Spq_neq0_uu_abs}). Consequently, the metric must be of the form
\begin{align}
\dd s^2 = & \left(1+\frac{\sR}{4(D-2)(D-3)}\,\delta_{mn}\,x^m x^n \right)^{-2}\,\delta_{pq}\, \dd x^p\,\dd x^q
\nonumber\\
&  -2\,\dd u\,\dd r + \left[\,\frac{4\Lambda_0-\sR}{(D-2)+2k \,(D-4)\sR}\,r^2+c(u)\,r+d(u, x)\,\right] \, \dd u^2 \,, \label{EGB-Kundt_constnat-curvature}
\end{align}
where $c(u)$ is an arbitrary function of retarded time $u$, while $d(u,x)$ satisfies the spatial Laplace equation (\ref{laplace}).

From the general form of the Weyl scalars (\ref{Psi2S})--(\ref{Psi4ij}) we find that the resulting spacetime is of \emph{algebraic type} II(bcd) because the ${(D-2)}$-dimensional transverse space is conformally flat Einstein space with the scalar curvature $\sR$ given by (\ref{sR_CC_GB}). The only nontrivial zero-boost-weight component (\ref{Psi2S}) reads
\begin{align}
\Psi_{2S} = \frac{D-3}{D-1}\left[\,b+\frac{\sR}{(D-2)(D-3)}\right]
    = \frac{(D-3)(4\Lambda_0-\sR)}{(D-1)\big[(D-2)+2k\, (D-4)\sR\big]}+\frac{\sR}{(D-1)(D-2)}
\,. \label{Psi2Sconstcurv}
\end{align}
Therefore, the class of solutions (\ref{EGB-Kundt_constnat-curvature}) with (\ref{laplace}) can be physically interpreted as \emph{exact type~II gravitational waves propagating on the type~D(bcd) background which is the  direct-product (anti-)Nariai universe}. Indeed, for ${D=4}$  we obtain ${\Psi_{2S} \equiv -2 Re(\Psi_2) = \frac{2}{3}\Lambda_0}$, i.e., ${\Psi_2=-\frac{1}{3}\Lambda_0}$ which fully agrees with the expressions in Sec.~7.2.1 of~\cite{GriffithsPodolsky:2009} and Secs.~18.6,~18.7 therein. We have thus found a generalization of the gravitational Kundt waves \cite{PodolskyOrtaggio:2003} to ${D>4}$ Einstein--Gauss--Bonnet gravity. These waves propagate in the higher-dimensional Nariai (${\sR>0}$) or anti-Nariai (${\sR<0}$) universe, identified previously in \cite{PodolskySvarc:2013a} (see, in particular, Sec.~11).

For the \emph{flat transverse space}, that is for ${\sR=0}$ and ${g_{pq}=\delta_{pq}}$ (and necessarily ${\Lambda_0=0}$), all the Gauss--Bonnet corrections in these solutions vanish, and we effectively deal with the Einstein theory. In fact, we end up in the subclass of VSI spacetimes (see \cite{CMPP:2004}) of the Weyl-type~N.

To illustrate the physical nature of the above solutions, we explicitly comment on the corresponding geodesic deviation (\ref{InvGeoDevExplicit_1_nat}), (\ref{InvGeoDevExplicit_i_nat}) of (transversally) \emph{static test observers}. In particular, the decomposition of the relative accelerations becomes
\begin{align}
\ddot Z^{(1)}=& \frac{2\left(\Lambda_0+2kH/D\right)}{(D-1)(D-2)}\,Z^{(1)}+\Psi_{2S} \,Z^{(1)}  +\frac{4\,k}{D-2}\,\cH_{ru} \,Z^{(1)} \,, \label{InvGeoDevExplicit_1_ConstC} \\
\ddot Z^{(i)}=& \frac{2\left(\Lambda_0+2k H/D\right)}{(D-1)(D-2)}\,Z^{(i)}-\frac{1}{D-2}\,\Psi_{2S} \,Z^{(i)}-\frac{1}{2}\Psi_{4^{ij}} \,Z^{(j)} \nonumber \\
&\hspace{10mm}  +\frac{k}{D-2}\left(-2m_i^pm_j^q\,\cH_{pq}\,Z^{(j)} +\big[(g_{uu}+2)\cH_{ru}+\cH_{uu}\big]Z^{(i)}\right), \label{InvGeoDevExplicit_i_ConstC}
\end{align}
where we used the relation ${\Psi_{2T^{(ij)}}=\frac{1}{D-2}\,\Psi_{2S}\,\delta_{ij}}$ since ${\tilde{\Psi}_{2T^{(ij)}}=0}$. By applying the field equations constraints, assuming a generic case with ${(D-2)+2k\, (D-4)\sR\neq0}$, the above quantities take the explicit form
\begin{align}
\Psi_{2S}&=\frac{D-3}{D-1}\Big[\,b+\frac{\sR}{(D-2)(D-3)}\,\Big] \,, \qquad \Psi_{4^{ij}}=-\tfrac{1}{2}m_i^pm_j^q\,d_{||pq} \,, \label{Psi4ijexpl}\\
H &= -\frac{1}{4}(D-4)\big[\sL+4b\sR\,\big] \,,  \qquad \cH_{pq} = \frac{2\,g_{pq}}{D(D-2)}\big[\sL-(D-4)b\sR\,\big] \,,\\
\cH_{ru}&=\tfrac{1}{2}g^{pq}\cH_{pq} \,, \qquad \cH_{uu}=-\frac{g_{uu}}{D}\big[\sL-(D-4)b\sR\,\big] \,,
\end{align}
with ${\sR}$ given by (\ref{sR_CC_GB}) and ${\sL=\frac{(D-4)(D-5)}{(D-2)(D-3)}\sR^2}$. The last term in equation (\ref{InvGeoDevExplicit_i_ConstC}) can thus be written as
\begin{equation}
\big[(g_{uu}+2)\cH_{ru}+\cH_{uu}\,\big] = \frac{2}{D}\big[\sL-(D-4)b\sR\,\big]=\frac{4(D-2)(\sR-2\Lambda_0)}{kD(D-5)\big[(D-2)+2k(D-4)\sR\,\big]} \,,
\end{equation}
which vanishes in ${D=4}$ corresponding to ${\sR=2\Lambda_0}$, see the constraint (\ref{CCS_GB_ru}). Combining all these explicit terms in (\ref{InvGeoDevExplicit_1_ConstC}), (\ref{InvGeoDevExplicit_i_ConstC}), we obtain a \emph{surprisingly simple result}
\begin{align}
\ddot Z^{(1)} &= b\,Z^{(1)} \,, \label{InvGeoDevExplicit_ConstC_fin1}\\
\ddot Z^{(i)} &= -\frac{1}{2}\Psi_{4^{ij}} \,Z^{(j)} \,, \label{InvGeoDevExplicit_ConstC_fini}
\end{align}
with $b$ given by (\ref{GB_c_g}) and ${\Psi_{4^{ij}}}$ given by (\ref{Psi4ijexpl}). The constant $b$ directly determines acceleration of the test particles along the \emph{longitudinal spatial direction}~$\bolde_{(1)}$, while ${\Psi_{4^{ij}}}$ (reflecting the non-trivial spacetime geometry via the corresponding covariant spatial derivatives ${d_{||pq}}$) causes \emph{symmetric and traceless deformations in the transverse directions}~$\bolde_{(i)}$ which represent exact \emph{Kundt--EGB gravitational waves}. Clearly, by setting ${d=0}$ in (\ref{EGB-Kundt_constnat-curvature}) the corresponding \emph{constant-curvature backgrounds} (without the waves) are obtained.

\newpage


\subsection{Einstein--Gauss--Bonnet {\it pp\,}-waves}

The class of {\it pp\,}-waves is invariantly defined as those geometries admitting a \emph{covariantly constant null vector field} \cite{Stephanietal:2003, GriffithsPodolsky:2009}. They thus necessary belong to the Kundt class with the privileged vector field ${\boldk=\partial_r}$. Moreover, the line element has to be $r$-independent, which implies
\begin{align}
g_{uu}\equiv d(u,x)  \label{d(ux)}
\end{align}
in the metric~(\ref{Kundt_non_gyr}), that is
\begin{equation}
\dd s^2 = g_{pq}(u,x)\, \dd x^p\,\dd x^q -2\,\dd u\,\dd r+d(u,x)\, \dd u^2 \,. \label{ppwavesmetric}
\end{equation}

In this case, the ${ru}$-component of the Einstein--Gauss--Bonnet field equations gives the same constraint (\ref{GB_ru}) as in the generic case, namely
\begin{align}
2\Lambda_0-\sR = k \,\Big(\sR^2_{klmn}-4\sR^2_{mn}+\sR^2 \Big) \,, \label{GB_pp_ru}
\end{align}
which can be used to eliminate the Gauss--Bonnet term of the transverse space ${g_{pq}}$ from the remaining equations.

Since we deal with the class with ${g_{uu,r}=0}$, the ${pq}$-component of the field equations (\ref{GB_pq_s}) becomes just an \emph{algebraic constraint on the admitted spatial curvature} in the form
\begin{align}
\sR_{pq}+2k \left(\sR_{pq}\sR-2\sR_{pmqn}\sR{}^{mn}+\sR_{pklm}\sR_q{}^{klm}-2\sR_{pm}\sR_q{}^m  \right) = 0 \,.
\label{ppconstraint}
\end{align}
The trace (\ref{GB_tr_s}) directly ties the spatial scalar curvature to the cosmological constant ${\Lambda_0}$ as
\begin{align}
\sR=4\Lambda_0 \,. \label{R=4lambda}
\end{align}
Let us emphasize that the case ${\Lambda_0 \ne 0}$ is \emph{not allowed in the Einstein theory}. Indeed, for ${k=0}$ the condition (\ref{GB_pp_ru}) requires ${\sR=2\Lambda_0}$, and in combination with (\ref{R=4lambda}) this necessarily leads to ${\Lambda_0=0=\sR}$.

With the above restrictions, the ${up}$-component of the field equations (\ref{GB_up_s}) now takes the form
\begin{align}
& g^{mn}\left[ \,g_{pn}-2k \left(2\sR_{pn}-\sR\, g_{pn}\right)\right]g^{kl}g_{k[m,u||l]} \nonumber\\
& \hspace{14.3mm} + 2k \left(2\sR{}^{kl}\,\delta_p^m - \sR_p{}^{kml}\right)g_{k[m,u||l]}=0 \,. \label{upComp_pp_waves}
\end{align}

Finally, the ${uu}$-component (\ref{GB_uu_s}) becomes
\begin{align}
& \hspace{-10.0mm} \big[-g^{mn} + 2k \left(2\sR^{mn}-\sR\, g^{mn}\right)\big]
\big(d_{||mn}+g_{mn,uu}-{\textstyle\frac{1}{2}}g^{pq}g_{pm,u}\,g_{qn,u}\big) \nonumber\\
& \hspace{26.0mm} +4k \left(g^{mo}g^{ns}-2\,g^{mn}g^{os}\right)g^{pq}g_{m[p,u||n]}\,g_{o[q,u||s]}=0 \,. \label{uuComp_pp_waves}
\end{align}

As an important explicit and non-trivial example of spacetimes satisfying all the above constraints, we consider a special case of the {\it pp\,}-wave geometries with \emph{constant-curvature transverse space} discussed in previous subsection~\ref{SubS:ConsTrSpace}. In particular, equation (\ref{ppconstraint}) with (\ref{const_curv_Riem}) is satisfied for (\ref{R=4lambda}) using the condition~(\ref{GB_pp_ru}). Moreover, these two constraints couple the spacetime dimension $D$ and the theory parameters $k$, $\Lambda_0$ as
\begin{equation}
\Lambda_0\,\bigg(8k\frac{(D-4)(D-5)}{(D-2)(D-3)}\,\Lambda_0 + 1\bigg) = 0 \,.
\end{equation}

There is an obvious solution ${\Lambda_0=0}$, equivalent to ${\sR=0}$, corresponding to \emph{flat transverse space}, as it appears in the Einstein theory, which represents \emph{planar} gravitational waves propagating on flat background. The \emph{new nontrivial class} with
\begin{equation}
\Lambda_0=-\frac{(D-2)(D-3)}{8k(D-4)(D-5)} \label{pp_waves_Lambda_nontrivial}
\end{equation}
is only allowed in the Einstein--Gauss--Bonnet theory. Since the transverse-space metric has to be $u$-independent, that is ${g_{pq}=g_{pq}(x)}$, equation (\ref{upComp_pp_waves}) is identically satisfied, and (\ref{uuComp_pp_waves}) greatly simplifies to
\begin{equation}
\frac{2}{D-5}\, \triangle d  =0 \,.
\end{equation}

\newpage

To summarize,
\begin{itemize}
\item
for ${\Lambda_0=0}$ we obtain the classic Weyl-type N solution
\begin{align}
\dd s^2 = \delta_{pq}\, \dd x^p\,\dd x^q -2\,\dd u\,\dd r + d(u, x) \, \dd u^2\,, \qquad \hbox{with} \qquad \delta^{ij}\,d_{,ij} =0 \,, \label{EGB-Kundt_pp_0_constnat-curvature}
\end{align}
\item
while in the non-Einsteinian case ${\Lambda_0\neq0}$ the spacetime becomes
\begin{align}
\dd s^2 = \bigg(1-\frac{\delta_{mn}\,x^m x^n}{8k(D-4)(D-5)}\bigg)^{-2}\,\delta_{pq}\, \dd x^p\,\dd x^q -2\,\dd u\,\dd r + d(u, x) \, \dd u^2\,, \qquad \hbox{with} \qquad \triangle d =0 \,, \label{EGB-Kundt_pp_constnat-curvature}
\end{align}
where the Laplace equation (\ref{laplace}) for ${d(u,x)}$ reflects the non-trivial transverse-space geometry, leading to the Weyl-type II(bcd) solutions.
\end{itemize}

In the classically forbidden case (\ref{EGB-Kundt_pp_constnat-curvature}), the geodesic deviation equations (\ref{InvGeoDevExplicit_1_nat}) and (\ref{InvGeoDevExplicit_i_nat}) take the form (\ref{InvGeoDevExplicit_1_ConstC}) and (\ref{InvGeoDevExplicit_i_ConstC}) where
\begin{align}
\Psi_{2S}&=\frac{4\Lambda_0}{(D-1)(D-2)} \,, \qquad \Psi_{4^{ij}}=-\tfrac{1}{2}m_i^pm_j^q\,d_{||pq} \,, \label{clen1}\\
H &= -\frac{1}{4}(D-4)\,\sL \,, \qquad \cH_{pq} = \frac{2\,g_{pq}}{D(D-2)}\,\sL \,, \\
\cH_{ru} &=\frac{1}{D}\,\sL \,, \qquad \cH_{uu}=-\frac{g_{uu}}{D}\,\sL \,, \qquad
\sL=16\Lambda_0^2\,\frac{(D-4)(D-5)}{(D-2)(D-3)}\,,
\end{align}
with $\Lambda_0$ given by (\ref{pp_waves_Lambda_nontrivial}). The additional Gauss--Bonnet contributions in equations (\ref{InvGeoDevExplicit_1_ConstC}) and (\ref{InvGeoDevExplicit_i_ConstC}) thus take the explicit form
\begin{align}
\Lambda_0+2kH/D &= -\frac{1}{4k}\frac{(D-2)^2(D-3)}{D(D-4)(D-5)} \,, \\
\sL &=\frac{1}{4k^2}\frac{(D-2)(D-3)}{(D-4)(D-5)} \,, \\
(g_{uu}+2)\cH_{ru}+\cH_{uu} &=\frac{1}{2k^2}\frac{(D-2)(D-3)}{D(D-4)(D-5)} \,. \label{clenN}
\end{align}

Based on the geodesic deviation, this class of vacuum solutions can be interpreted as \emph{exact gravitational {\it pp\,}-waves}, represented by the ${\Psi_{4^{ij}}}$ components, which propagate on \emph{Weyl-type D(bcd) background} whose isotropic influence is encoded in the term ${\Lambda_0+2kH/D}$. This background causes the Newtonian behavior encoded in the $\Psi_{2S}$ scalar, combined with the Gauss--Bonnet contributions ${\cH_{pq}}$ and ${\cH_{ru}}$, respectively, and with the additional transverse effect given by the term ${(g_{uu}+2)\cH_{ru}+\cH_{uu}}$. Interestingly, these different curvature components perfectly combine. Summing up all the explicit terms (\ref{clen1})--(\ref{clenN}), the complete form of the geodesic deviation equations (\ref{InvGeoDevExplicit_1_ConstC}) and (\ref{InvGeoDevExplicit_i_ConstC}) becomes \emph{extremely simple}, namely
\begin{align}
\ddot Z^{(1)} &= 0 \,, \label{InvGeoDevExplicit_ConstCpp_fin1}\\
\ddot Z^{(i)} &= -\frac{1}{2}\Psi_{4^{ij}} \,Z^{(j)} \,. \label{InvGeoDevExplicit_ConstCpp_fini}
\end{align}
This is a special case of (\ref{InvGeoDevExplicit_ConstC_fin1}), (\ref{InvGeoDevExplicit_ConstC_fini}) when ${b=0 \Leftrightarrow \sR=4\Lambda_0}$.

It describes \emph{purely transverse and traceless tidal deformations}, without any longitudinal effects. In Einstein's theory, these would be interpreted as the typical effect of \emph{Weyl-type N gravitational waves} propagating in \emph{flat} Minkowski space. Surprisingly, in the context of Einstein--Gauss--Bonnet theory the \emph{Weyl tensor remains of algebraic type II(bcd)} with
\begin{equation}
\Psi_{2S}=\frac{4\Lambda_0}{(D-1)(D-2)} = -\frac{(D-3)}{2k(D-1)(D-4)(D-5)}  \ne 0\,. \label{Psi2bacground form EGBppwaves}
\end{equation}
Such transverse \emph{EGB~{\it pp\,}-waves propagate on non-flat constant-curvature (anti-)Nariai background} given by the metric (\ref{EGB-Kundt_pp_constnat-curvature}) with ${d\equiv 0}$ (which is (\ref{EGB-Kundt_constnat-curvature}) when ${c=0=d}$ and  ${\sR=4\Lambda_0\equiv-\frac{(D-2)(D-3)}{2k(D-4)(D-5)}}$, see (\ref{pp_waves_Lambda_nontrivial})).
The wave amplitudes are geometrically encoded in the scalars ${\Psi_{4^{ij}}\equiv -\frac{1}{2}m_i^pm_j^q\,d_{||pq} }$. The specific imprint of the non-flat background is thus only encoded in the \emph{non-trivial covariant derivatives} on its constant-curvature transverse wave-front, entering $\Psi_{4^{ij}}$ via $d_{||pq}$. This is the only way how to distinguish the two globally distinct types of gravitational waves in the LISA-type gravitational wave detector.


\section{Conclusions}
\label{Sec:Conclusions}

Assuming the family of spacetime manifolds which admit a non-twisting, non-expanding and shear-free null geodesic congruence, constituting the famous Kundt class of geometries, we derived, discussed and analyzed the corresponding exact vacuum solutions to the Einstein--Gauss--Bonnet theory in an arbitrary dimension ${D\geq 5}$. Our only additional restriction was the absence of the so called gyratonic terms ${g_{up}}$, leading to the initial metric ansatz (\ref{Kundt_non_gyr}).

Starting with the quantities characterizing the spacetime curvature of (\ref{Kundt_non_gyr}), summarized in Appendices~\ref{appendixA} and~\ref{appendixB}, in Section~\ref{Sec:FEqs} we derived the fully general form (\ref{GB_ru})--(\ref{GB_uu_s}) of the field equations. In the subsections we distinguished and presented three distinct subclasses defined by the key tensorial quantity $\Spq$ and its trace~$Q$ given by (\ref{Spq}) and (\ref{SpqTRACE}), respectively.

In the subsequent Section~\ref{Sec:AlgStr} we introduced a natural null frame and analyzed the algebraic structure of the Weyl and also the traceless Ricci tensor in terms of the corresponding frame projections and their irreducible parts. Within the metric (\ref{Kundt_non_gyr}), these tensors are algebraically special. In particular, all positive boost-weight components are vanishing, see (\ref{Psi2S})--(\ref{Psi4ij}) and (\ref{Phi00nat})--(\ref{Phi22nat}). Moreover, further specializations enter via specific field equations constraints.

To better understand the physical nature of the resulting solution, we presented the invariant form of the geodesic deviation equation in Section~\ref{Sec:GeodDev}. Its crucial ingredient, the Riemann curvature tensor, was decomposed to its traceless Weyl part and the Ricci tensor and scalar. The corresponding frame projections were expressed in terms of the scalar quantities introduced in previous Section~\ref{Sec:AlgStr}. Moreover, the Ricci contributions were further re-expressed in terms of the Gauss--Bonnet part of the field equations (\ref{RicciFieldEqs}) to explicitly identify the effects of such a theory on relative motion of free test particles, detectable in principle by the LISA-type gravitational wave detectors. The result is given by equations (\ref{InvGeoDevExplicit_1}), (\ref{InvGeoDevExplicit_i}), and in the subsequent paragraph discussing the specific deformations of the geodesic congruence associated with a time-like observer.

Finally, in Section~\ref{Sec:Subclasses} we discussed two most important representatives of the Kundt family, namely the Kundt gravitational waves and backgrounds with ${(D-2)}$-dimensional transverse space being of a constant curvature, and the complete family of {\it pp\,}-waves defined as geometries admitting a covariantly constant null vector field. In the first case the resulting line element is (\ref{EGB-Kundt_constnat-curvature}), while for the {\it pp\,}-waves we obtained two possible types of explicit metrics (\ref{EGB-Kundt_pp_0_constnat-curvature}) and (\ref{EGB-Kundt_pp_constnat-curvature}).


\section*{Acknowledgements}

This work was supported by the Czech Science Foundation Grant No.~GA\v{C}R 20-05421S. It is our informal contribution to broad investigations within the LISA Consortium, whose Associate members we are, concerning the fundamental physical properties of gravitational waves in alternative theories of gravity.


\newpage

\appendix

\section{Curvature tensors for the Kundt geometry}
\label{appendixA}

For the $D$-dimensional Kundt geometry (\ref{Kundt_non_gyr}) with ${g_{up}=0}$, the non-vanishing Christoffel symbols are
\begin{align}
& \Gamma^r_{ru} = -\frac{1}{2}g_{uu,r} \,, &
& \Gamma^r_{uu} = \frac{1}{2}g_{uu}\,g_{uu,r}-\frac{1}{2}g_{uu,u} \,, &
& \Gamma^r_{up} = -\frac{1}{2}g_{uu,p} \,, &
& \Gamma^r_{pq} = \frac{1}{2}g_{pq,u} \,, \nonumber \\
& \Gamma^u_{uu} = \frac{1}{2}g_{uu,r} \,, &
& \Gamma^m_{uu} = -\frac{1}{2}g^{mn}\,g_{uu,n} \,, &
& \Gamma^m_{up} = \frac{1}{2}g^{mn}\,g_{np,u} \,, &
& \Gamma^m_{pq} = \,^{S}\Gamma^m_{pq} \,,
\end{align}
where ${\,^{S}\Gamma^m_{pq}}$ denotes the Christoffel symbols of the spatial metric ${g_{pq}}$ on the transverse ${(D-2)}$-dimensional space with coordinates ${x^p}$. The corresponding non-vanishing Riemann curvature tensor components read
\begin{align}
R_{ruru} &= -\frac{1}{2}g_{uu,rr} \,, \qquad
R_{mpnq} = \sR_{mpnq}  \,, \qquad
R_{ruup} = \frac{1}{2}\,g_{uu,rp} \,, \qquad
R_{upmq} = g_{p[m,u||q]} \,, \nonumber \\
R_{upuq} &= -\frac{1}{2}g_{uu||pq}-\frac{1}{2}g_{pq,uu}+\frac{1}{4}g_{uu,r}g_{pq,u}+\frac{1}{4}g^{mn}g_{mp,u}g_{nq,u} \,,\label{Riem_upuq}
\end{align}
with ${\,_{||}}$ denoting the covariant derivative on the transverse space, i.e., with respect to the connection ${\,^{S}\Gamma^m_{pq}}$. Also ${\sR_{mpnq}}$ stands for the transverse-space Riemann tensor. Finally, the non-zero Ricci tensor components~are
\begin{align}
R_{ru} &= -\frac{1}{2}g_{uu,rr} \,, \qquad\quad
R_{pq} = \sR_{pq} \,, \qquad\quad
R_{up} = -\frac{1}{2}g_{uu,rp}+g^{mn}g_{m[p,u||n]} \,, \nonumber \\
R_{uu} &= \frac{1}{2}g_{uu}\,g_{uu,rr}+\frac{1}{4}g^{mn}g_{mn,u}\,g_{uu,r}-\frac{1}{2}g^{mn}g_{mn,uu} -\frac{1}{2}g^{mn}\,g_{uu||mn}+\frac{1}{4}g^{mn}g^{pq}g_{pm,u}\,g_{qn,u} \,, \label{Ricci_all}
\end{align}
and the Ricci scalar curvature is
\begin{equation}\label{Ricci_scalar}
R = \sR+g_{uu,rr} \,,
\end{equation}
with ${\sR_{pq}\equiv g^{mn}\sR_{mpnq}}$ and ${\sR\equiv g^{pq}\sR_{pq}}$, respectively.


\section{Specific quadratic terms for the Kundt geometry}
\label{appendixB}

To evaluate the Gauss--Bonnet term $L_{GB}$ (\ref{GB_term}) for the geometries (\ref{Kundt_non_gyr}), we have to express squares of the Riemann and Ricci tensors, and the scalar curvature. The result is
\begin{align}
R^2_{cdef} &= \sR^2_{klmn}+(g_{uu,rr})^2 \,,&
R^2_{cd} &= \sR^2_{mn}+\frac{1}{2}\,(g_{uu,rr})^2 \,,&
R^2 &= \left(\sR+g_{uu,rr}\right)^2 .&
\end{align}
Moreover, the non-vanishing curvature tensors contractions appearing in $H_{ab}$ in the field equations (\ref{GB_Field_eqns})~are
\begin{itemize}
\item ${ru}$-component
\begin{align}
R_{rc}\,R_u{}^c &= -\frac{1}{4}(g_{uu,rr})^2 \,,&
R_{rcud}\,R^{cd} &= -\frac{1}{4}(g_{uu,rr})^2 \,,&
R_{rcde}\,R_u{}^{cde} &= -\frac{1}{2}(g_{uu,rr})^2 \,,&
\end{align}

\item ${pq}$-component
\begin{align}
R_{pc}\,{R_q}^c &= \sR_{pm}\sR_q{}^m \,,&
R_{pcqd}\,R^{cd} &= \sR_{pmqn}\sR^{mn} \,,&
R_{pcde}\,R_q{}^{cde} &= \sR_{pklm}\sR_q{}^{klm} \,,&
\end{align}

\item ${up}$-component
\begin{align}
R_{uc}\,R_p{}^c &= \frac{1}{2}g_{uu,rr}\Big(-\frac{1}{2}g_{uu,rp}+g^{mn}g_{m[p,u||n]}\Big)+g^{mn}\sR_{mp}\Big(-\frac{1}{2}g_{uu,rn}+g^{kl}g_{k[n,u||l]}\Big) \,, \nonumber \\
R_{ucpd}\,R^{cd} &= -\frac{1}{4}g_{uu,rr}\,g_{uu,rp}+g_{m[p,u||n]}\sR^{mn} \,, \nonumber \\
R_{ucde}\,R_p{}^{cde} &= -\frac{1}{2}g_{uu,rr}\,g_{uu,rp}+g_{k[l,u||m]}\sR_p{}^{klm} \,,
\end{align}

\item ${uu}$-component
\begin{align}
R_{uc}\,R_u{}^c &= g^{pq}\Big(-\frac{1}{2}g_{uu,rp}+g^{mn}\,g_{m[p,u||n]}\Big)\Big(-\frac{1}{2}g_{uu,rq}+g^{os}\,g_{o[q,u||s]}\Big)\nonumber\\
& \hspace{0.0mm} + \frac{1}{4}g_{uu,rr}\left(g_{uu}\,g_{uu,rr}+g^{mn}g_{mn,u}\,g_{uu,r}-2g^{mn}g_{mn,uu}-2g^{mn} g_{uu||mn}+g^{mn}g^{pq}g_{mp,u}\,g_{nq,u}\right) ,  \nonumber \\
R_{ucud}\,R^{cd} &= \frac{1}{8}g_{uu,rr}\left(2g_{uu}\,g_{uu,rr}-g^{mn}g_{mn,u}\,g_{uu,r}+2g^{mn}g_{mn,uu}+2g^{mn} g_{uu||mn}-g^{mn}g^{pq}g_{mp,u}\,g_{nq,u}\right) \nonumber \\
& \hspace{0.0mm} + g^{pq}g_{uu,rp}\Big(-\frac{1}{2}g_{uu,rq}+g^{mn}g_{m[q,u||n]}\Big) \nonumber \\
& \hspace{0.0mm} +\frac{1}{4}\sR^{pq}\left(-2g_{uu||pq}-2g_{pq,uu}+g_{uu,r}\,g_{pq,u}+g^{mn}g_{mp,u}\,g_{nq,u}\right) , \nonumber \\
R_{ucde}\,R_u{}^{cde} &=\frac{1}{2}g_{uu}(g_{uu,rr})^2-\frac{1}{2}g^{pq}g_{uu,rp}\,g_{uu,rq}+g^{os}g^{mn}g^{pq}g_{o[m,u||p]}\,g_{s[n,u||q]} \,.
\end{align}
\end{itemize}
Using these relations, the Gauss--Bonnet contribution ${H_{ab}}$ to the field equations, see (\ref{GB_Field_eqns}) and (\ref{GB_FE_contrib}),  explicitly becomes
\begin{align}
H_{ru} =& \frac{1}{4} \sL \label{Hru} \,, \\
H_{pq} =& \sH_{pq}+g_{uu,rr}\Big(\sR_{pq}-\frac{1}{2}\sR \,g_{pq}\Big)\label{Hpq} \,, \\
H_{up} =& {\sR_p}^ng_{uu,rn}-\frac{1}{2}\sR \,g_{uu,rp}-2\Big(\sR^{mn}-\frac{1}{2}\sR \,g^{mn}\Big)g_{m[p,u||n]}+g_{k[l,u||m]}\Big({\sR_p}^{klm}-2{\sR_p}^lg^{km}\Big) \label{Hup} \,, \\
H_{uu} =& \Big(\sR^{pq}-\frac{1}{2}\sR\, g^{pq}\Big)
\left(g_{uu||pq}+g_{pq,uu}-\frac{1}{2}g_{uu,r}\,g_{pq,u}-\frac{1}{2}g^{kl}g_{kp,u}\,g_{lq,u}\right) \nonumber \\
&+\left(g^{mo}g^{ns}-2\,g^{mn}g^{os}\right)g^{pq}g_{m[p,u||n]}\,g_{o[q,u||s]} -\frac{1}{4}g_{uu}\sL\label{Huu} \,,
\end{align}
where ${\sL\equiv \sR^2_{cdef}-4\,\sR^2_{cd}+\sR^2 }$ is the Gauss--Bonnet term of the transverse-space geometry. For the trace of $H_{ab}$ we obtain
\begin{equation}
H=-\frac{1}{4}(D-4)\,L_{GB} = -\frac{1}{4}(D-4)\left(\sL+2\sR\, g_{uu,rr}\right). \label{HExpl}
\end{equation}


\end{document}